\documentclass[twocolumn,tighten]{aastex62}

\usepackage{natbib}
\usepackage{graphicx}		
\usepackage{url}
\usepackage{epstopdf}
\usepackage{color, xcolor}
\usepackage{multirow}
\usepackage{ragged2e}
\usepackage{hyperref}
\usepackage{amsmath} 
\usepackage{mathtools}
\usepackage{booktabs}
\usepackage{comment}
\usepackage{float}
\usepackage{sidecap}
\usepackage{wrapfig}
\usepackage{ulem}
\usepackage{footmisc}
\usepackage{array}
\usepackage{capt-of}

\DeclareGraphicsExtensions{.pdf,.png,.jpeg}

\defcitealias{Psaltis_2018}{P18} 
\defcitealias{Johnson_2018}{J18} 
\defcitealias{Issaoun_2019}{I19} 

\definecolor{orange}{rgb}{1,0.5,0}

\hypersetup{colorlinks=true, citecolor={blue}}

\newcommand\sgra{Sgr\,A$^{*}$\,}

\newcommand{\rev}[1]{#1}
\newcommand{\revnew}[1]{#1}

\accepted{December 8, 2021}
\submitjournal{ApJ}

\begin{document}

\title{The intrinsic structure of Sagittarius\,A* at 1.3\,cm and 7\,mm}
\shorttitle{The intrinsic structure of \sgra at 1.3\,cm and 7\,mm}

\correspondingauthor{G.-Y. Zhao}
\email{gyzhao@iaa.es}

\author[0000-0001-6083-7521]{Ilje Cho} 
\affiliation{Korea Astronomy and Space Science Institute, Daedeok-daero 776, Yuseong-gu, Daejeon 34055, Republic of Korea} 
\affiliation{University of Science and Technology, Gajeong-ro 217, Yuseong-gu, Daejeon 34113, Republic of Korea}
\affiliation{Instituto de Astrof\'{\i}sica de Andaluc\'{\i}a - CSIC, Glorieta de la Astronom\'{\i}a s/n, E-18008 Granada, Spain}
\shortauthors{Cho et al.}

\author[0000-0002-4417-1659]{Guang-Yao Zhao} 
\affiliation{Instituto de Astrof\'{\i}sica de Andaluc\'{\i}a - CSIC, Glorieta de la Astronom\'{\i}a s/n, E-18008 Granada, Spain}
\affiliation{Korea Astronomy and Space Science Institute, Daedeok-daero 776, Yuseong-gu, Daejeon 34055, Republic of Korea} 

\author[0000-0001-8527-0496]{Tomohisa Kawashima}
\affiliation{Institute for Cosmic Ray Research, The University of Tokyo, 5-1-5 Kashiwanoha, Kashiwa, Chiba 277-8582, Japan}

\author[0000-0002-2709-7338]{Motoki Kino}
\affiliation{National Astronomical Observatory of Japan, 2-21-1 Osawa, Mitaka, Tokyo 181-8588, Japan}
\affiliation{Kogakuin University of Technology \& Engineering, Academic Support Center, 
2665-1 Nakano, Hachioji, Tokyo 192-0015, Japan}

\author[0000-0002-9475-4254]{Kazunori Akiyama}
\affiliation{Massachusetts Institute of Technology Haystack Observatory, 99 Millstone Road, Westford, MA 01886, USA}
\affiliation{Black Hole Initiative at Harvard University, 20 Garden Street, Cambridge, MA 02138, USA}
\affiliation{National Astronomical Observatory of Japan, 2-21-1 Osawa, Mitaka, Tokyo 181-8588, Japan}

\author[0000-0002-4120-3029]{Michael D. Johnson}
\affiliation{Center for Astrophysics $\vert$ Harvard \& Smithsonian, 60 Garden Street, Cambridge, MA 02138, USA}
\affiliation{Black Hole Initiative at Harvard University, 20 Garden Street, Cambridge, MA 02138, USA}

\author[0000-0002-5297-921X]{Sara Issaoun}
\affiliation{Department of Astrophysics, Institute for Mathematics, Astrophysics and Particle Physics (IMAPP), Radboud University, P.O. Box 9010, 6500 GL Nijmegen, The Netherlands}

\author[0000-0003-1364-3761]{Kotaro Moriyama}
\affiliation{Mizusawa VLBI Observatory, National Astronomical Observatory of Japan, 2-12 Hoshigaoka, Mizusawa, Oshu, Iwate 023-0861, Japan}
\affiliation{Massachusetts Institute of Technology Haystack Observatory, 99 Millstone Road, Westford, MA 01886, USA}

\author[0000-0003-4407-9868]{Xiaopeng Cheng}
\affiliation{Korea Astronomy and Space Science Institute, Daedeok-daero 776, Yuseong-gu, Daejeon 34055, Republic of Korea} 

\author[0000-0001-6993-1696]{Juan-Carlos Algaba}
\affiliation{Department of Physics, Faculty of Science, University of Malaya, 50603 Kuala Lumpur, Malaysia}

\author[0000-0001-7003-8643]{Taehyun Jung}
\affiliation{Korea Astronomy and Space Science Institute, Daedeok-daero 776, Yuseong-gu, Daejeon 34055, Republic of Korea} 
\affiliation{University of Science and Technology, Gajeong-ro 217, Yuseong-gu, Daejeon 34113, Republic of Korea}

\author[0000-0002-4148-8378]{Bong Won Sohn}
\affiliation{Korea Astronomy and Space Science Institute, Daedeok-daero 776, Yuseong-gu, Daejeon 34055, Republic of Korea}
\affiliation{University of Science and Technology, Gajeong-ro 217, Yuseong-gu, Daejeon 34113, Republic of Korea}
\affiliation{Department of Astronomy, Yonsei University, Yonsei-ro 50, Seodaemun-gu, Seoul 03722, Republic of Korea}

\author[0000-0002-4892-9586]{Thomas P. Krichbaum}
\affiliation{Max-Planck-Institut f\"ur Radioastronomie, Auf dem H\"ugel 69, D-53121 Bonn, Germany}

\author[0000-0002-8635-4242]{Maciek Wielgus}
\affiliation{Center for Astrophysics $\vert$ Harvard \& Smithsonian, 60 Garden Street, Cambridge, MA 02138, USA}
\affiliation{Black Hole Initiative at Harvard University, 20 Garden Street, Cambridge, MA 02138, USA}

\author[0000-0001-6906-772X]{Kazuhiro Hada}
\affiliation{Mizusawa VLBI Observatory, National Astronomical Observatory of Japan, 2-12 Hoshigaoka, Mizusawa, Oshu, Iwate 023-0861, Japan}
\affiliation{Department of Astronomical Science, The Graduate University for Advanced Studies (SOKENDAI), 2-21-1 Osawa, Mitaka, Tokyo 181-8588, Japan}

\author[0000-0002-7692-7967]{Ru-Sen Lu}
\affiliation{Shanghai Astronomical Observatory, Chinese Academy of Sciences, 80 Nandan Road, Shanghai 200030, China}
\affiliation{Key Laboratory of Radio Astronomy, Chinese Academy of Sciences, Nanjing 210008, China}
\affiliation{Max-Planck-Institut f\"ur Radioastronomie, Auf dem H\"ugel 69, D-53121 Bonn, Germany}

\author[0000-0001-6311-4345]{Yuzhu Cui}
\affiliation{Mizusawa VLBI Observatory, National Astronomical Observatory of Japan, 2-12 Hoshigaoka, Mizusawa, Oshu, Iwate 023-0861, Japan}
\affiliation{Department of Astronomical Science, The Graduate University for Advanced Studies (SOKENDAI), 2-21-1 Osawa, Mitaka, Tokyo 181-8588, Japan}

\author[0000-0001-7719-274X]{Satoko Sawada-Satoh}
\affiliation{The Research Institute for Time Studies, Yamaguchi University, 1677-1 Yoshida, Yamaguchi, Yamaguchi 753-8511, Japan}

\author[0000-0003-3540-8746]{Zhiqiang Shen}
\affiliation{Shanghai Astronomical Observatory, Chinese Academy of Sciences, 80 Nandan Road, Shanghai 200030, China}
\affiliation{Key Laboratory of Radio Astronomy, Chinese Academy of Sciences, Nanjing 210008, China}

\author[0000-0001-6558-9053]{Jongho Park} 
\altaffiliation{EACOA fellow}
\affiliation{Institute of Astronomy and Astrophysics, Academia Sinica, 11F of Astronomy-Mathematics Building, AS/NTU No. 1, Sec. 4, Roosevelt Rd, Taipei 10617, Taiwan, R.O.C.}
\affiliation{Department of Physics and Astronomy, Seoul National University, Gwanak-gu, Seoul 08826, Republic of Korea}

\author[0000-0001-7369-3539]{Wu Jiang}
\affiliation{Shanghai Astronomical Observatory, Chinese Academy of Sciences, 80 Nandan Road, Shanghai 200030, China}
\affiliation{Key Laboratory of Radio Astronomy, Chinese Academy of Sciences, Nanjing 210008, China}


\author[0000-0002-7322-6436]{Hyunwook Ro}         
\affiliation{Department of Astronomy, Yonsei University, Yonsei-ro 50, Seodaemun-gu, Seoul 03722, Republic of Korea}
\affiliation{Korea Astronomy and Space Science Institute, Daedeok-daero 776, Yuseong-gu, Daejeon 34055, Republic of Korea}

\author{Kunwoo Yi}         
\affiliation{Department of Physics and Astronomy, Seoul National University, Gwanak-gu, Seoul 08826, Republic of Korea}

\author[0000-0003-3823-7954]{Kiyoaki Wajima}      
\affiliation{Korea Astronomy and Space Science Institute, Daedeok-daero 776, Yuseong-gu, Daejeon 34055, Republic of Korea}

\author[0000-0003-2147-0290]{Jee Won Lee}        
\affiliation{Korea Astronomy and Space Science Institute, Daedeok-daero 776, Yuseong-gu, Daejeon 34055, Republic of Korea}

\author{Jeffrey Hodgson}     
\affiliation{Department of Physics and Astronomy, Sejong University, 209 Neungdong-ro, Gwangjin-gu, Seoul, South Korea}

\author[0000-0003-0236-0600]{Fumie Tazaki}
\affiliation{Mizusawa VLBI Observatory, National Astronomical Observatory of Japan, 2-12 Hoshigaoka, Mizusawa, Oshu, Iwate 023-0861, Japan}

\author[0000-0003-4058-9000]{Mareki Honma}
\affiliation{Mizusawa VLBI Observatory, National Astronomical Observatory of Japan, 2-12 Hoshigaoka, Mizusawa, Oshu, Iwate 023-0861, Japan}
\affiliation{Department of Astronomical Science, The Graduate University for Advanced Studies (SOKENDAI), 2-21-1 Osawa, Mitaka, Tokyo 181-8588, Japan}

\author[0000-0002-8169-3579]{Kotaro Niinuma}      
\affiliation{Graduate School of Sciences and Technology for Innovation, Yamaguchi University, 1677-1 Yoshida, Yamaguchi, Yamaguchi 753-8511, Japan}
\affiliation{The Research Institute for Time Studies, Yamaguchi University, Yoshida 1677-1, Yamaguchi, Yamaguchi 753-8511, Japan}

\author[0000-0003-0465-1559]{Sascha Trippe}
\affiliation{Department of Physics and Astronomy, Seoul National University, Gwanak-gu, Seoul 08826, Republic of Korea}
\affiliation{SNU Astronomy Research Center, Seoul National University, Gwanak-gu, Seoul 08826, Republic of Korea}

\author[0000-0003-4341-0029]{Tao An}              
\affiliation{Shanghai Astronomical Observatory, Chinese Academy of Sciences, 80 Nandan Road, Shanghai 200030, China}

\author{Yingkang Zhang}     
\affiliation{Shanghai Astronomical Observatory, Chinese Academy of Sciences, 80 Nandan Road, Shanghai 200030, China}

\author{Jeong Ae Lee} 
\affiliation{Korea Astronomy and Space Science Institute, Daedeok-daero 776, Yuseong-gu, Daejeon 34055, Republic of Korea}


\author{Se-Jin Oh}          
\affiliation{Korea Astronomy and Space Science Institute, Daedeok-daero 776, Yuseong-gu, Daejeon 34055, Republic of Korea}

\author[0000-0003-1157-4109]{Do-Young Byun}
\affiliation{Korea Astronomy and Space Science Institute, Daedeok-daero 776, Yuseong-gu, Daejeon 34055, Republic of Korea}
\affiliation{University of Science and Technology, Gajeong-ro 217, Yuseong-gu, Daejeon 34113, Republic of Korea}

\author[0000-0002-6269-594X]{Sang-Sung Lee}
\affiliation{Korea Astronomy and Space Science Institute, Daedeok-daero 776, Yuseong-gu, Daejeon 34055, Republic of Korea}
\affiliation{University of Science and Technology, Gajeong-ro 217, Yuseong-gu, Daejeon 34113, Republic of Korea}

\author[0000-0001-8229-7183]{Jae-Young Kim}
\affiliation{Korea Astronomy and Space Science Institute, Daedeok-daero 776, Yuseong-gu, Daejeon 34055, Republic of Korea}

\author{Junghwan Oh}         
\affiliation{Korea Astronomy and Space Science Institute, Daedeok-daero 776, Yuseong-gu, Daejeon 34055, Republic of Korea}

\author[0000-0002-3723-3372]{Shoko Koyama}
\affiliation{Niigata University, 8050 Ikarashi 2-no-cho, Nishi-ku, Niigata 950-2181, Japan}
\affiliation{Institute of Astronomy and Astrophysics, Academia Sinica, 11F of Astronomy-Mathematics Building, AS/NTU No. 1, Sec. 4, Roosevelt Rd, Taipei 10617, Taiwan, R.O.C.}

\author{Keiichi Asada}
\affiliation{Institute of Astronomy and Astrophysics, Academia Sinica, 11F of Astronomy-Mathematics Building, AS/NTU No. 1, Sec. 4, Roosevelt Rd, Taipei 10617, Taiwan, R.O.C.}

\author{Xuezheng Wang}  
\affiliation{Shanghai Astronomical Observatory, Chinese Academy of Sciences, 80 Nandan Road, Shanghai 200030, China}

\author{Lang Cui}           
\affiliation{Xinjiang Astronomical Observatory, Chinese Academy of Sciences, Urumqi 830011, China}
\affiliation{Key Laboratory of Radio Astronomy, Chinese Academy of Sciences, Nanjing 210008, China}

\author[0000-0002-9043-6048]{Yoshiaki Hagiwara}   
\affiliation{Toyo University, 5-28-20 Hakusan, Bunkyo-ku, Tokyo 112-8606, Japan}

\author[0000-0001-6081-2420]{Masanori Nakamura}
\affiliation{National Institute of Technology, Hachinohe College, Yubinbango Aomori Prefecture Hachinohe Oaza Tamonoki character Ueno flat 16-1, 039-1192, Japan}
\affiliation{Institute of Astronomy and Astrophysics, Academia Sinica, 11F of Astronomy-Mathematics Building, AS/NTU No. 1, Sec. 4, Roosevelt Rd, Taipei 10617, Taiwan, R.O.C.}

\author{Mieko Takamura}      
\affiliation{National Astronomical Observatory of Japan, 2-21-1 Osawa, Mitaka, Tokyo 181-8588, Japan}

\author[0000-0003-1659-095X]{Tomoya Hirota}      
\affiliation{National Astronomical Observatory of Japan, 2-21-1 Osawa, Mitaka, Tokyo 181-8588, Japan}

\author[0000-0002-6033-5000]{Koichiro Sugiyama}  
\affiliation{National Astronomical Research Institute of Thailand (Public Organization), 260 Moo 4, T. Donkaew, A. Maerim, Chiangmai, 50180, Thailand}

\author[0000-0002-7776-3159]{Noriyuki Kawaguchi} 
\affiliation{National Astronomical Observatory of Japan, 2-21-1 Osawa, Mitaka, Tokyo 181-8588, Japan}

\author{Hideyuki Kobayashi} 
\affiliation{Mizusawa VLBI Observatory, National Astronomical Observatory of Japan, 2-12 Hoshigaoka, Mizusawa, Oshu, Iwate 023-0861, Japan} 

\author{Tomoaki Oyama}
\affiliation{Mizusawa VLBI Observatory, National Astronomical Observatory of Japan, 2-12 Hoshigaoka, Mizusawa, Oshu, Iwate 023-0861, Japan} 

\author[0000-0001-5615-5464]{Yoshinori Yonekura}
\affiliation{Center for Astronomy, Ibaraki University, 2-1-1 Bunkyo, Mito, Ibaraki 310-8512, Japan} 

\author[0000-0002-1229-0426]{Jongsoo Kim}
\affiliation{Korea Astronomy and Space Science Institute, Daedeok-daero 776, Yuseong-gu, Daejeon 34055, Republic of Korea}

\author{Ju-Yeon Hwang}
\affiliation{Korea Astronomy and Space Science Institute, Daedeok-daero 776, Yuseong-gu, Daejeon 34055, Republic of Korea}

\author{Dong-Kyu Jung}
\affiliation{Korea Astronomy and Space Science Institute, Daedeok-daero 776, Yuseong-gu, Daejeon 34055, Republic of Korea}

\author{Hyo-Ryoung Kim}
\affiliation{Korea Astronomy and Space Science Institute, Daedeok-daero 776, Yuseong-gu, Daejeon 34055, Republic of Korea}

\author{Jeong-Sook Kim}
\affiliation{\revnew{Basic Science Research Institute, Chungbuk National University, Chungdae-ro 1, Seowon-Gu, Cheongju, Chungbuk 28644, Korea}}

\author{Chung-Sik Oh}
\affiliation{Korea Astronomy and Space Science Institute, Daedeok-daero 776, Yuseong-gu, Daejeon 34055, Republic of Korea}

\author{Duk-Gyoo Roh}
\affiliation{Korea Astronomy and Space Science Institute, Daedeok-daero 776, Yuseong-gu, Daejeon 34055, Republic of Korea}

\author{Jae-Hwan Yeom}
\affiliation{Korea Astronomy and Space Science Institute, Daedeok-daero 776, Yuseong-gu, Daejeon 34055, Republic of Korea}

\author{Bo Xia} 
\affiliation{Shanghai Astronomical Observatory, Chinese Academy of Sciences, 80 Nandan Road, Shanghai 200030, China}
\affiliation{Key Laboratory of Radio Astronomy, Chinese Academy of Sciences, Nanjing 210008, China}

\author{Weiye Zhong} 
\affiliation{Shanghai Astronomical Observatory, Chinese Academy of Sciences, 80 Nandan Road, Shanghai 200030, China}
\affiliation{Key Laboratory of Radio Astronomy, Chinese Academy of Sciences, Nanjing 210008, China}

\author{Bin Li} 
\affiliation{Shanghai Astronomical Observatory, Chinese Academy of Sciences, 80 Nandan Road, Shanghai 200030, China}
\affiliation{Key Laboratory of Radio Astronomy, Chinese Academy of Sciences, Nanjing 210008, China}

\author{Rongbing Zhao} 
\affiliation{Shanghai Astronomical Observatory, Chinese Academy of Sciences, 80 Nandan Road, Shanghai 200030, China}
\affiliation{Key Laboratory of Radio Astronomy, Chinese Academy of Sciences, Nanjing 210008, China}

\author{Jinqing Wang} 
\affiliation{Shanghai Astronomical Observatory, Chinese Academy of Sciences, 80 Nandan Road, Shanghai 200030, China}
\affiliation{Key Laboratory of Radio Astronomy, Chinese Academy of Sciences, Nanjing 210008, China}

\author{Qinghui Liu} 
\affiliation{Shanghai Astronomical Observatory, Chinese Academy of Sciences, 80 Nandan Road, Shanghai 200030, China}
\affiliation{Key Laboratory of Radio Astronomy, Chinese Academy of Sciences, Nanjing 210008, China}

\author{Zhong Chen} 
\affiliation{Shanghai Astronomical Observatory, Chinese Academy of Sciences, 80 Nandan Road, Shanghai 200030, China}
\affiliation{Key Laboratory of Radio Astronomy, Chinese Academy of Sciences, Nanjing 210008, China}


\begin{abstract}
Sagittarius A* (\sgra), the Galactic Center supermassive black hole (SMBH), is one of the best targets to resolve the innermost region of SMBH with very long baseline interferometry (VLBI). 
In this study, we have carried out observations toward \sgra at 1.349\,cm (22.223\,GHz) and 6.950\,mm (43.135\,GHz) with the East Asian VLBI Network, as a part of the multi-wavelength campaign of the Event Horizon Telescope (EHT) in 2017 April. 
To mitigate scattering effects, the physically motivated scattering kernel model from \citet{Psaltis_2018} and the scattering parameters from \citet{Johnson_2018} have been applied. 
As a result, a single, symmetric Gaussian model well describes the intrinsic structure of \sgra at both wavelengths. 
From closure amplitudes, the major-axis sizes are $\sim704\,\pm\,102\,\mu$as (axial ratio$\sim1.19^{+0.24}_{-0.19}$) and $\sim300\,\pm\,25\,\mu$as (axial ratio$\sim1.28\,\pm\,0.2$) at 1.349\,cm and 6.95\,mm respectively. 
Together with a quasi-simultaneous observation at 3.5\,mm (86\,GHz) by \citet{Issaoun_2019}, we show that the intrinsic size \revnew{scales with} observing wavelength \revnew{as a} power-law\revnew{, with an} index $\sim1.2\pm0.2$. 
Our results also provide estimates of the size and compact flux density at 1.3\,mm, which can be incorporated into the analysis of the EHT observations.
In terms of the origin of radio emission, we have compared the intrinsic structures with the accretion flow scenario, especially the radiatively inefficient accretion flow based on \revnew{the} Keplerian shell model. With this, we show that a nonthermal electron \rev{population} is necessary to reproduce the source sizes. 
\\
\end{abstract}


\section{Introduction} 
\label{sec:intro}

The supermassive black hole (SMBH) at our Galactic Center, Sagittarius A* (\sgra), is the closest SMBH to the Earth, with a mass $\sim4.1\times10^6\,M_{\odot}$ \citep{Ghez_2008, Genzel_2010}. 
Thanks to its proximity, $\sim$8.1\,kpc \citep[e.g.,][]{Gravity_2019}, 
\sgra has the largest angular size of the Schwarzschild radius ($R_{s}$) among all black holes, $\sim10\,\mu$as. 
Therefore it is one of the best laboratories to explore mass accretion onto SMBHs, and it is the other main target besides M\,87 for Event Horizon Telescope (EHT) to resolve the black hole shadow \citep{ehtc2019a,ehtc2019b,ehtc2019c,ehtc2019d,ehtc2019e,ehtc2019f}. 

In very long baseline interferometry (VLBI) observations at centimeter\,(cm) wavelengths, the structure of \sgra is dominated by scatter broadening caused by the ionized interstellar scattering medium (ISM; see, e.g., \citealt{Rickett_1990, Narayan_1992}) \rev{which is located at $\sim5.8\,$kpc from Earth \citep{Bower_2014a}}. 
As a result, the observed sizes are proportional to $\lambda^2$ where $\lambda$ is the observing wavelength 
\citep{Davies_1976,vanLangevelde_1992,Lo_1998,Bower_2004,Shen_2005,Bower_2006,Johnson_2018}. 
The shape has been found to be an asymmetric Gaussian, elongated towards the east-west (i.e., stronger angular broadening; \citealt{Lo_1985,Lo_1998,Alberdi_1993,Frail_1994,Bower_Backer_1998}). 
While the major and minor axis sizes are sub-dominant to scattering at $\gtrsim1\,$cm, there are marginal detections of the measured sizes larger than the $\lambda^2$ scaling at millimeter\,(mm) wavelengths which indicate the intrinsic size of \sgra \citep{Krichbaum_1998,Lo_1998,Doeleman_2001,Lu_2011}. 
Especially at 1.3\,mm, \revnew{the} persistent asymmetric structure of \sgra has been also suggested from the proto-EHT observations \citep{Fish_2016, Lu_2018}. 

In addition to the angular broadening by the diffractive scattering, refractive scattering effects introduce substructure in the image which \revnew{is} caused by the density irregularities in the ionized ISM. 
This substructure appears as ``refractive noise'' in interferometric visibilities \citep{Goodman_1989, Narayan_1989, Johnson_Gwinn_2015, Johnson_Narayan_2016}, and it was discovered at 1.3\,cm observations of \sgra by \citet{Gwinn_2014}. 
At 7 and 3\,mm, non-zero closure phases have been detected which can be interpreted as either the imprint of the intrinsic asymmetry of \sgra or the refractive substructure \citep{Rauch_2016, Ortiz-Leon_2016,Brinkerink_2016,Brinkerink_2019}. 
To better constrain the scattering effects, recently, \citet{Psaltis_2018} (hereafter \citetalias{Psaltis_2018}) suggested a physically motivated scattering model for anisotropic scattering. 
They introduced the magnetohydro dynamics (MHD) turbulence with a finite inner scale and a wandering transverse magnetic field direction to their simplified model and derived analytic models of expected observational properties, such as the scatter broadening and refractive scintillation. 
With this model, \citet{Johnson_2018} (hereafter \citetalias{Johnson_2018}) tightly constrained the scattering parameters such as power-law index of the phase structure function of the scattering screen, $\alpha$, and a finite inner scale of interstellar turbulence, $r_{\rm in}$, using multi-wavelength observational data. 
After that, \citet{Issaoun_2019} (hereafter \citetalias{Issaoun_2019}) showed that the scattering model from \citetalias{Johnson_2018} provided comparable refractive noise to their observations at 3.5\,mm using the Global Millimeter VLBI Array (GMVA) in concert with the phased Atacama Large Millimeter/submillimeter Array (ALMA) in 2017 April.

In this study, we analyze the properties of \sgra at 1.349\,cm (22.223\,GHz; hereafter 1.3\,cm and 22\,GHz) and 6.950\,mm (43.135\,GHz; hereafter 7\,mm and 43\,GHz) with the East Asian VLBI Network (EAVN; \citealt{Hagiwara_2015, Wajima_2016, An_2018, Cui_2021}) observations in 2017 April, by mitigating the scattering effects based on the studies from \citetalias{Psaltis_2018}, \citetalias{Johnson_2018}, and \citetalias{Issaoun_2019}. 
The observations have been carried out within two days from the EHT (1.3\,mm) and GMVA+ALMA (3.5\,mm; \citetalias{Issaoun_2019}) campaign, providing a range of quasi-simultaneous multi-wavelength observations, which is crucial for constraining the physical parameters. To this end, we report the flux densities and intrinsic structures of \sgra at both wavelengths. 

\rev{
Especially the structure of \sgra can provide an important hint of its emission model. There is a long-debate \revnew{about} the dominant emission source of \sgra, whether an accretion flow or a jet. While there are many theoretical predictions \citep[e.g.,][]{Moscibrodzka_2014} and indirect \revnew{evidence} \citep[e.g.,][]{Brinkerink_2015} of a possible jet from \sgra, a jet-like structure has not been resolved yet. 
So we first find the most probable model of an accretion flow scenario, if it can solely explain the observed results in which circumstances. 
For this purpose, we have compared the measured intrinsic structure of \sgra with radiatively inefficient accretion flow (RIAF) based on \revnew{the} Keplerian shell model \citep{Falcke_2000,Broderick_2006,Pu_2016,Kawashima_2019} and found that the thermal-nonthermal electron population is necessary to reproduce the observational results. 
\revnew{A} similar comparison has been carried out with spectral energy distribution (SED; e.g., \citealt{Ozel_2000, Yuan_2003}) and with theoretical structure at 230\,GHz \citep{Chael_2017, Mao_2017}, but has not been investigated with the structure at lower frequency range (i.e., 22 and 43\,GHz). 
Note that our analysis is based on a recent study of \revnew{the} electron energy spectrum in an accretion flow \citep{Broderick_2011, Pu_2016}, so it does not rule out the possible jet scenario which is beyond the scope of this paper. 
}

\rev{In this paper, }
we begin with the outline of our observations and data processing in \autoref{sec:obs_calibration}. 
Next, in \autoref{sec:imfit}, the imaging and model--fitting procedures are described. 
In \autoref{sec:results}, the overall results including the wavelength-dependence and its implication on the EHT imaging are presented. 
In \autoref{sec:nonthermal_model}, the importance of nonthermal electron components for an accretion flow model to explain our results is discussed. 
Lastly, we summarize our findings in \autoref{sec:summary}. 
\\

\section{Observations and Data reduction} \label{sec:obs_calibration}

\begin{table}[t]
\centering
\caption{EAVN and the quasi-simultaneous observations}
\begin{tabular}{ccccc}
\hline
\hline
Array & 
Date (2017) & 
$\lambda$\,(cm) & 
Reference \\
\hline
EAVN & Apr~03 & 1.349 & This work \\
EAVN & Apr~04 & 0.695 & This work \\
GMVA+ALMA & Apr~03 & 0.35 & \citetalias{Issaoun_2019} \\
EHT & Apr~05-11 & 0.13 & ... \\
\hline
\end{tabular}
\\ \vspace{0.3cm}
\raggedright{\textbf{Note. }
From left to right, the observing array, date, wavelength, and references. } 
\label{tab:obs}
\end{table}

\begin{figure*}[t]
\centering 
\includegraphics[width=\columnwidth]{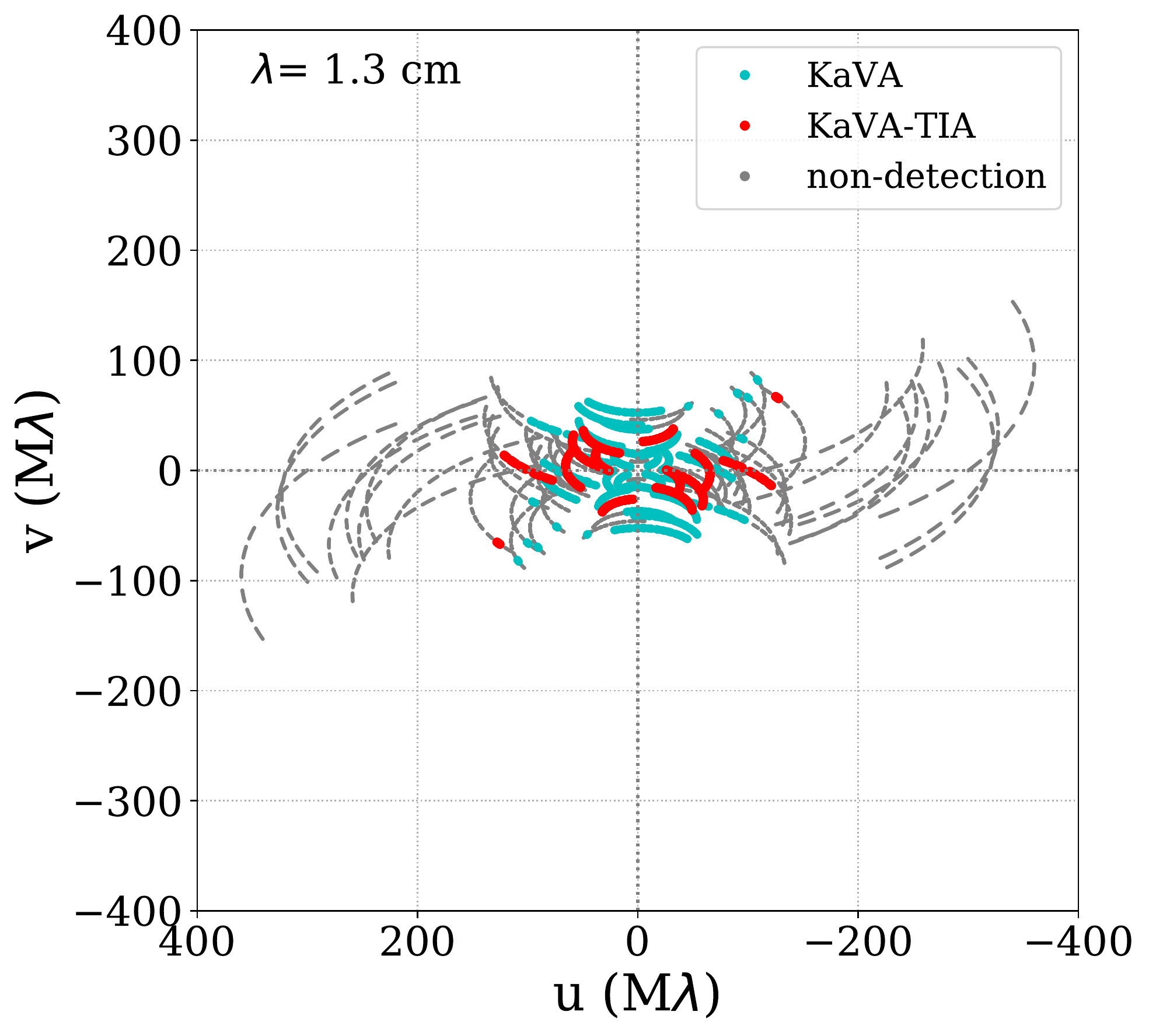}
\includegraphics[width=\columnwidth]{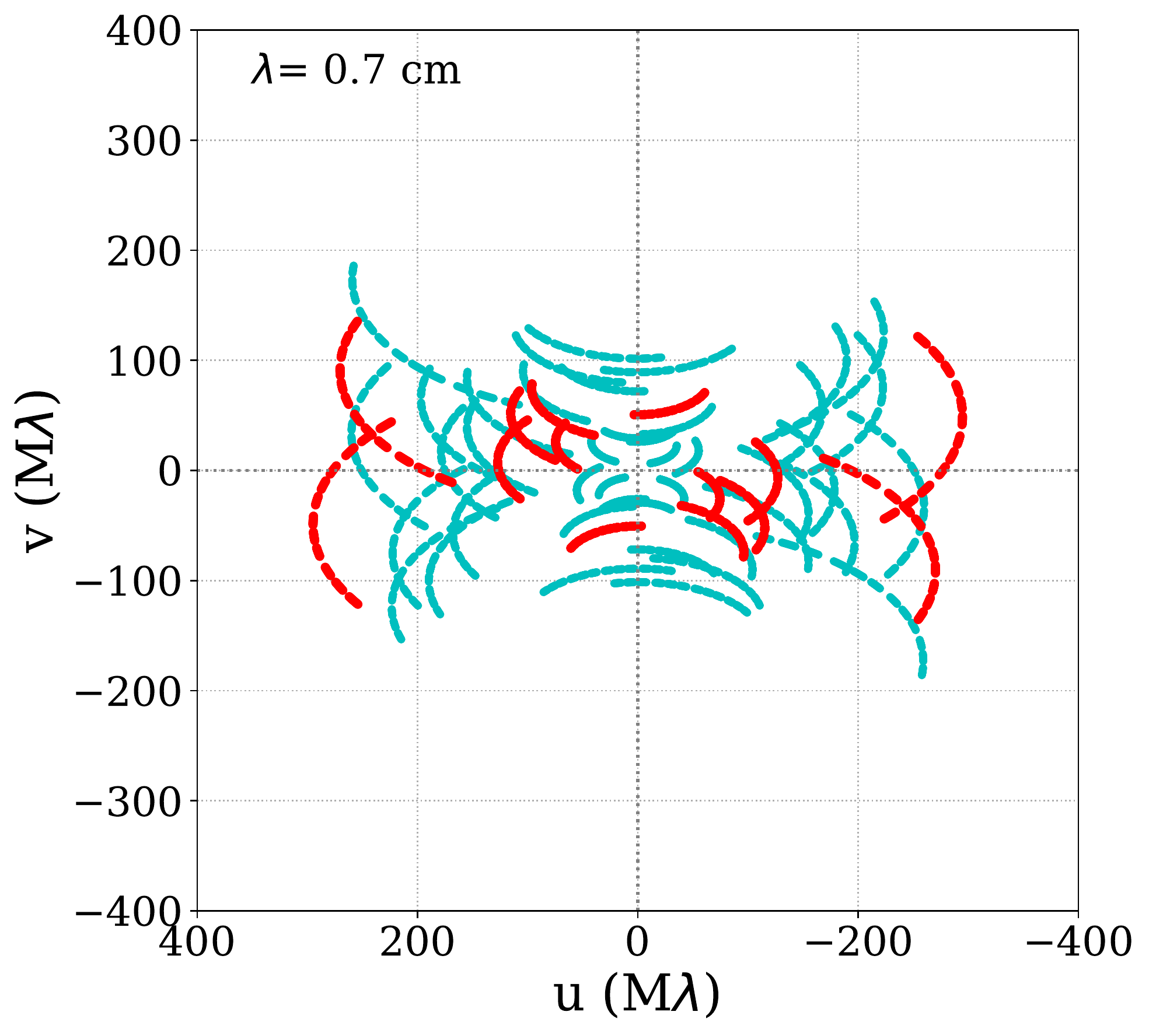}
\caption{
$u-v$ coverage of EAVN observations at 1.3\,cm (left) and 0.7\,cm (right). 
The KaVA observations with successful detection (S/N$>5$; cyan) and flagged points (gray) are shown. 
TIA provides the longest baselines (red). Each point has been averaged within 30\,second interval. 
}
\label{fig:uvplt}
\end{figure*}

The EAVN consists of the KaVA 
(KVN\footnote{Korean VLBI Network, which consists of three 21\,m telescopes in Korea: Yonsei (KYS), Ulsan (KUS), and Tamna (KTN)} 
and VERA\footnote{VLBI Exploration of Radio Astrometry, which consists of four 20\,m telescopes in Japan: Mizusawa (MIZ), Iriki (IRK), Ogasawara (OGA), and Ishigakijima (ISG)} Array; e.g., \citealt{Niinuma_2015, Hada_2017, Park_2019}) and additional East-Asian telescopes (e.g., Tianma-65m, Nanshan-26m, and Hitachi-32m telescopes; \citealt{Cui_2021}). 
As a part of the KaVA/EAVN Large Program \citep[e.g.,][]{Kino_2015} which intensively monitors \sgra within $\lesssim$ a month time interval, several EAVN observations at 1.3\,cm and 7\,mm were carried out in 2017 April. 
In this study, we present the results of two observations (April 3 and 4, 2017) close to the other global VLBI campaigns: GMVA+ALMA at 3.5\,mm (April 3, 2017) and EHT at 1.3\,mm (April 5$-$11, 2017; see \autoref{tab:obs}). 
The EAVN data are recorded with 256\,MHz (32\,MHz$\,\times\,$8\,channels) total bandwidth. 
The on-source time for \sgra and a calibrator, NRAO\,530, \revnew{is} $\sim$160 and $\sim$30\,minutes respectively. 

The KaVA and Tianma-65m (TIA) participated in the observations at both 1.3\,cm and 7\,mm (\autoref{fig:uvplt}). 
At 1.3\,cm, the Nanshan-26m (Urumqi, UR), and Hitachi-32m (HT) telescopes additionally joined. 
However, UR provides the baseline length $\sim 100-380\,{\rm M}\lambda$ where is already dominated by the refractive scattering noises (see, \autoref{sec:imfit}) so the fringes were not detected. 
The HT had a severe problem with its antenna gain so it was flagged out. 

The data were calibrated using NRAO Astronomical Imaging Processing System (AIPS; \citealt{Greisen_2003}). 
The cross-power spectra were first normalized using the auto-correlation power spectra ({\tt ACCOR} in AIPS), and a multiplicative correction factor of 1.3 was applied to all data to correct the quantization loss in the Daejeon hardware correlator \citep{Lee_2015}. 
For the amplitude calibration, we used the system temperature and antenna gain curve information (a-priori calibration method; {\tt APCAL} in AIPS) for 1.3\,cm data. As for the 7\,mm data, the SiO maser lines from OH\,0.55$-$0.06 and VX\,Sgr were instead used which provided a better gain correction (template spectrum method; {\tt ACFIT} in AIPS). 
\citet{Cho_2017} have confirmed that the template spectrum method can derive \revnew{a} more realistic antenna gain curve as a function of the elevation than a-priori calibration method so that better accuracy of the amplitude calibration can be achieved.  
Note that it is also possible to use the H$_{2}$O maser line at 1.3\,cm but the nearby maser source, Sgr\,B2, has an extended maser spot distribution up to $\sim120\,$arcseconds which is much broader than the beam size of TIA (e.g., full width half maximum, FWHM, $\sim44\,$arcseconds at 1.3\,cm) so that it may provide relatively worse results \citep{Cho_2017}. 
The phase calibration was implemented with three steps: 1) the ionospheric effect and parallactic angle corrections \footnote{Since VERA stations have equatorial mounts, the parallactic angle corrections are applied to KVN.} 
({\tt VLBATECR} and {\tt VLBAPANG} in AIPS), 
2) instrumental phase and delay offset correction using the best fringe fit solutions of a fringe tracer, and 3) the global fringe search ({\tt FRING} in AIPS). 
Note that the {\tt VLBATECR} was not applied for 7\,mm data since the ionospheric effect was less severe. The correction of {\tt VLBAPANG} was also small for KaVA data so that the visibility phases were mostly consistent before and after the calibration. 
As a result, the phase delay and fringe frequency solutions were successfully obtained with \revnew{a} signal-to-noise ratio (S/N) $>$\,5. 
The bandpass calibration was done in two steps: 1) for the amplitudes using the total power spectra, and 2) for both amplitudes and phases using the cross power spectra of \sgra. 

After the calibrations, the visibilities at lower elevation $<5^\circ$ at either telescope of a baseline were flagged. 
To correct the residual amplitudes offset across the frequency channels (8\,channels), both the multi-channel data and channel-averaged data were used. 
First, a Gaussian model was obtained from the channel-averaged data. 
The multi-channel data were then self-calibrated with the Gaussian model and averaged across the entire bandwidth. 
The frequency-averaged visibilities were coherently time averaged over 30 seconds, and the outliers were flagged. 
\\

\section{Imaging and model fitting} \label{sec:imfit}

\begin{figure*}[t]
\centering 
\includegraphics[width=0.3\linewidth]{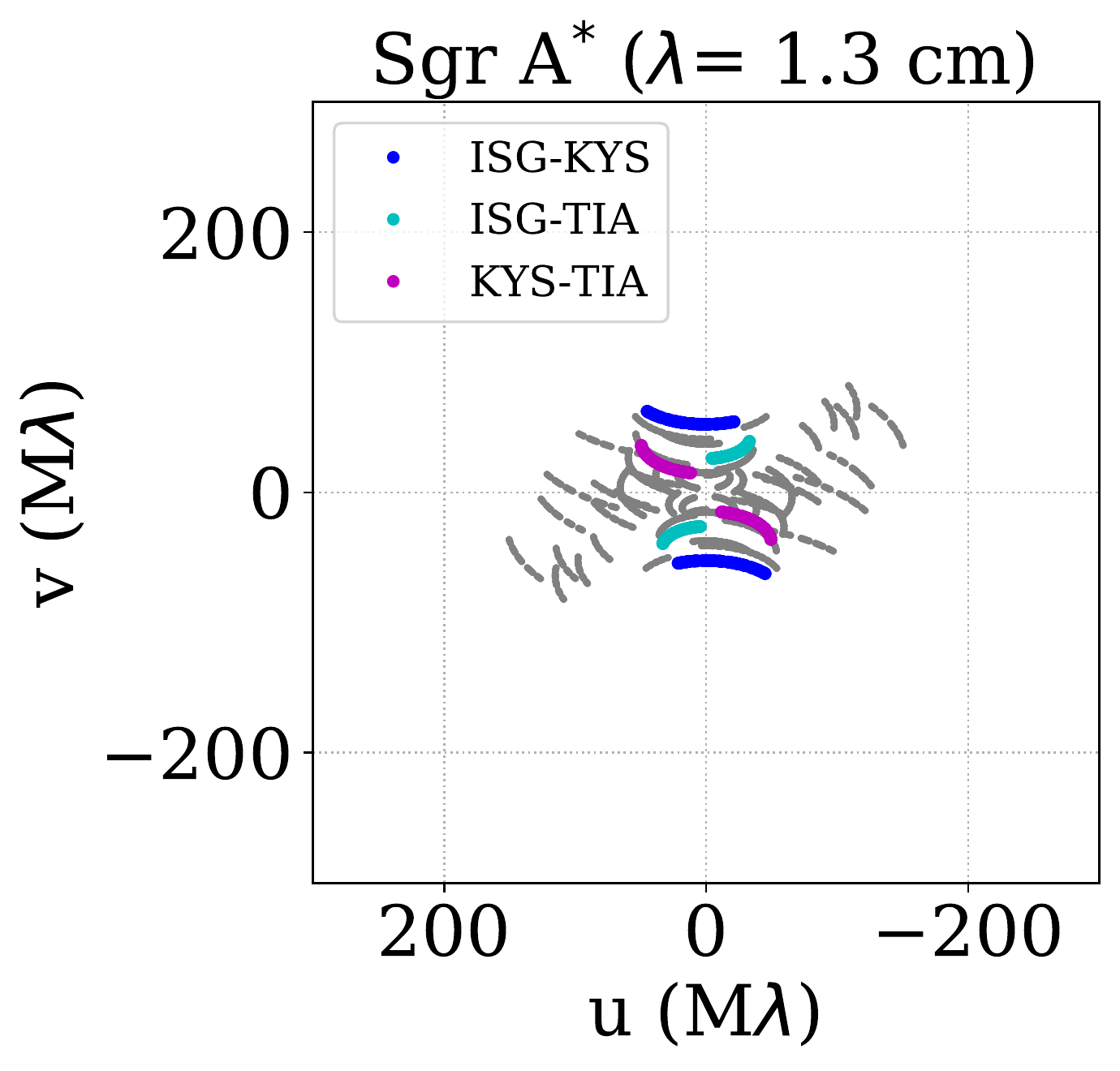}
\includegraphics[width=0.5\linewidth]{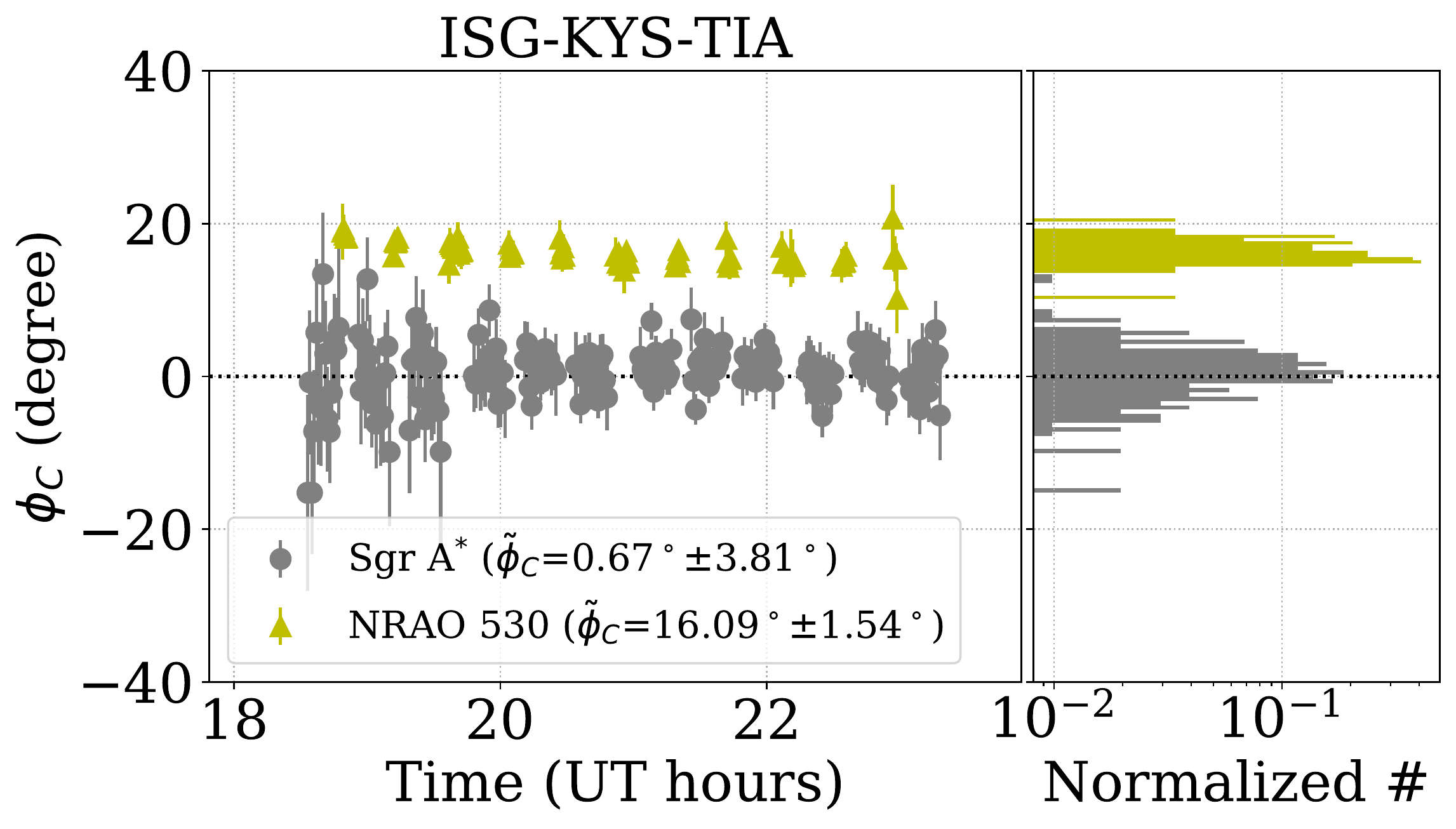} \\
\includegraphics[width=0.3\linewidth]{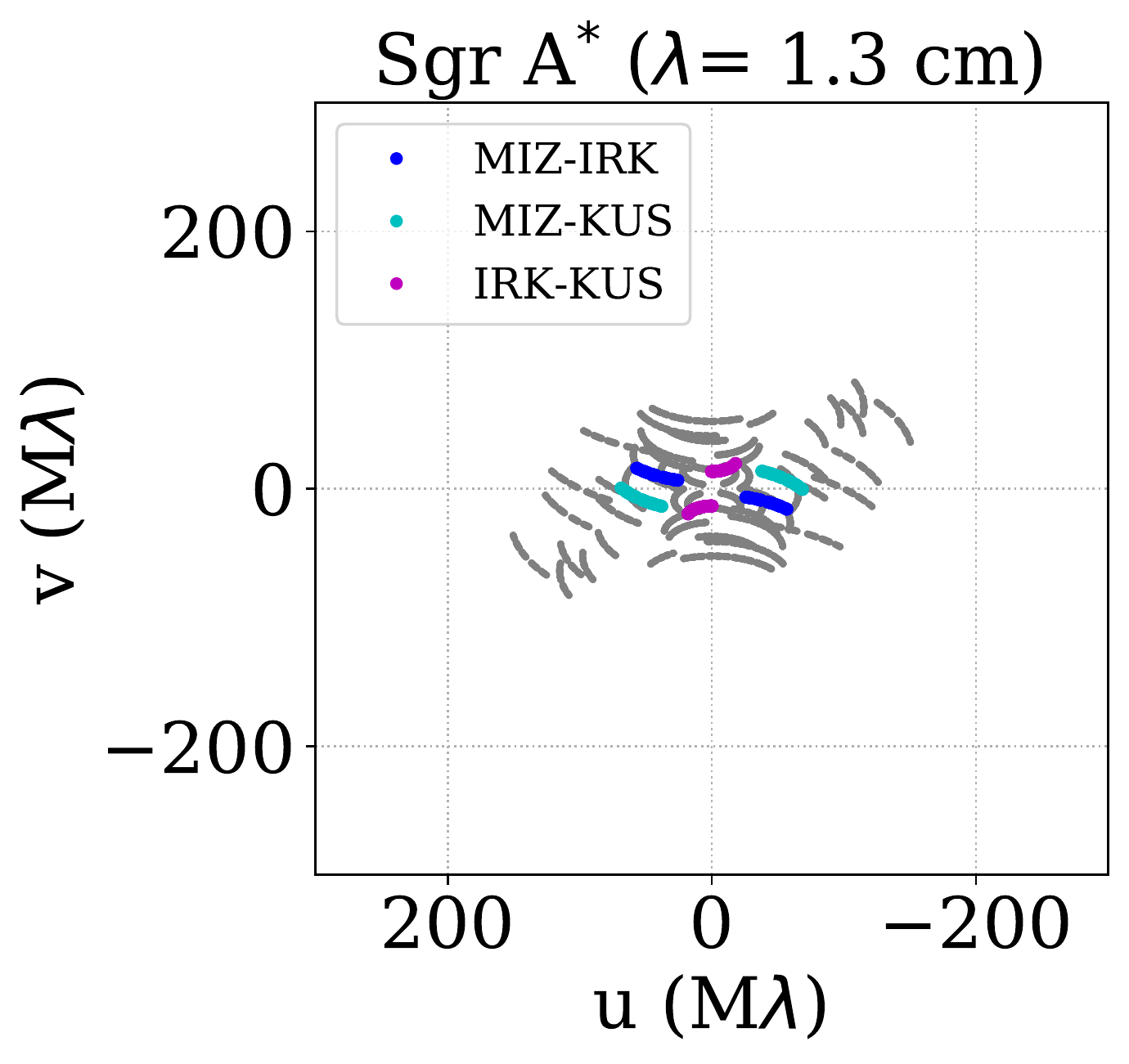} 
\includegraphics[width=0.5\linewidth]{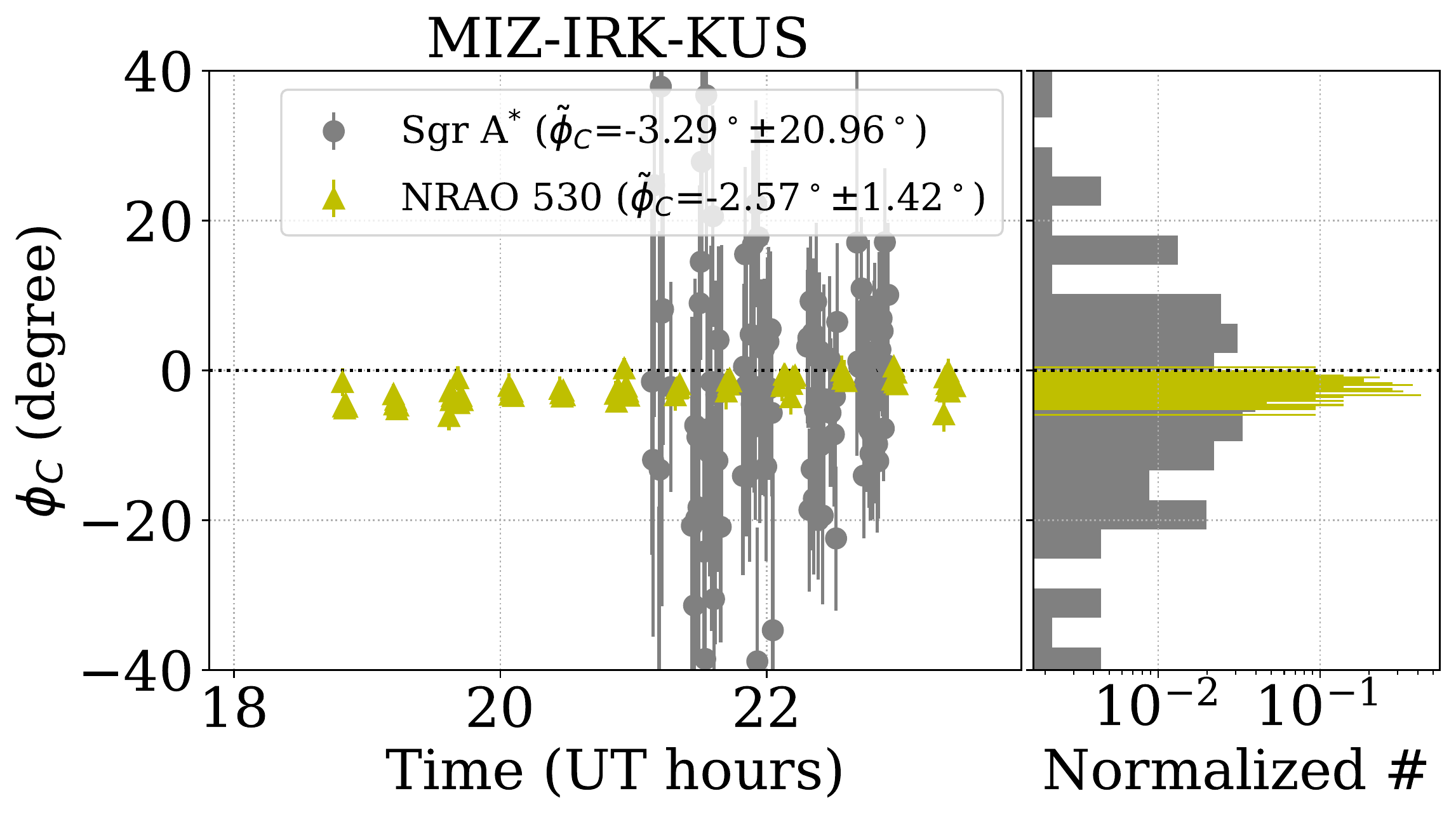} \\
\includegraphics[width=0.3\linewidth]{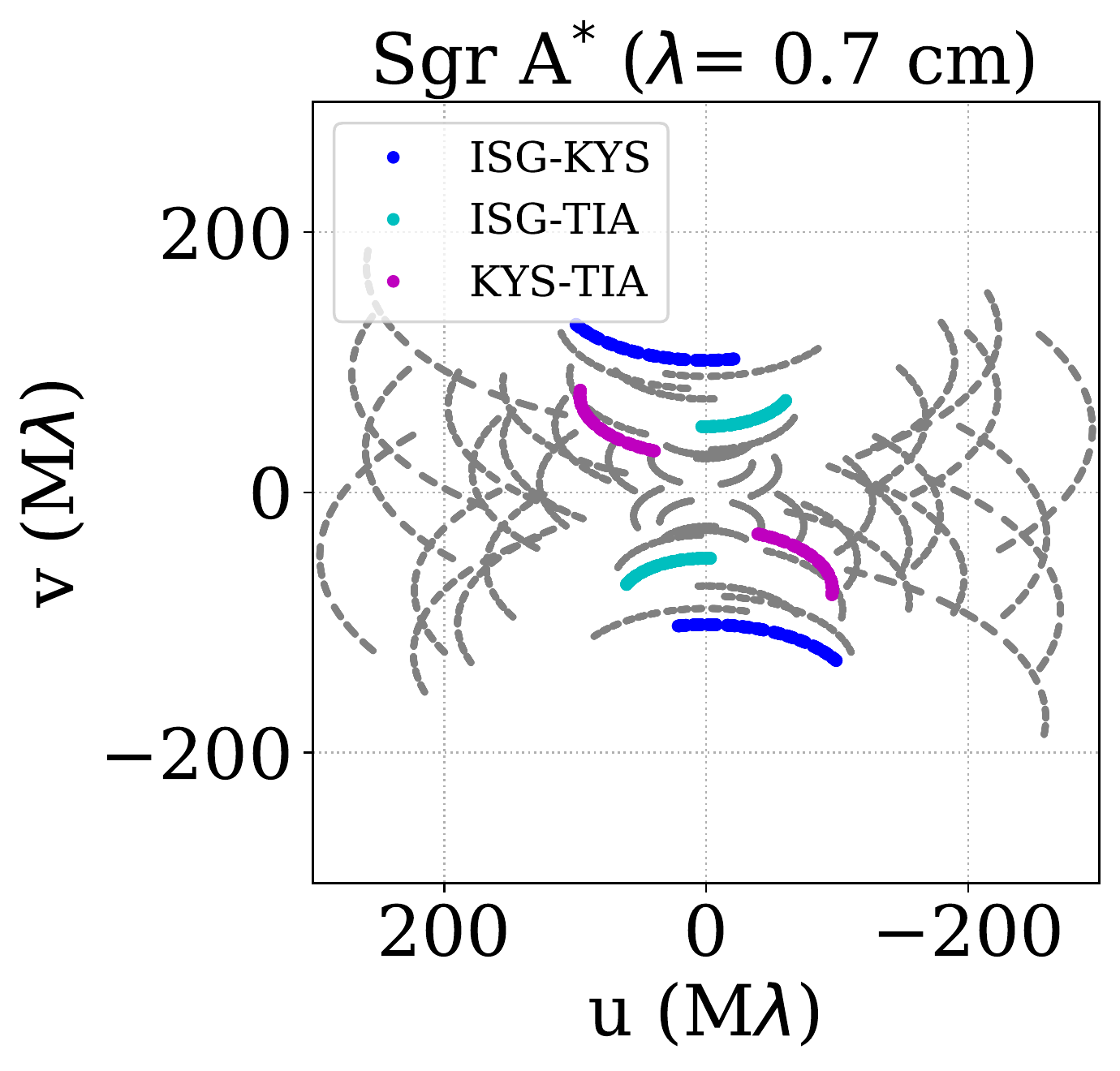} 
\includegraphics[width=0.5\linewidth]{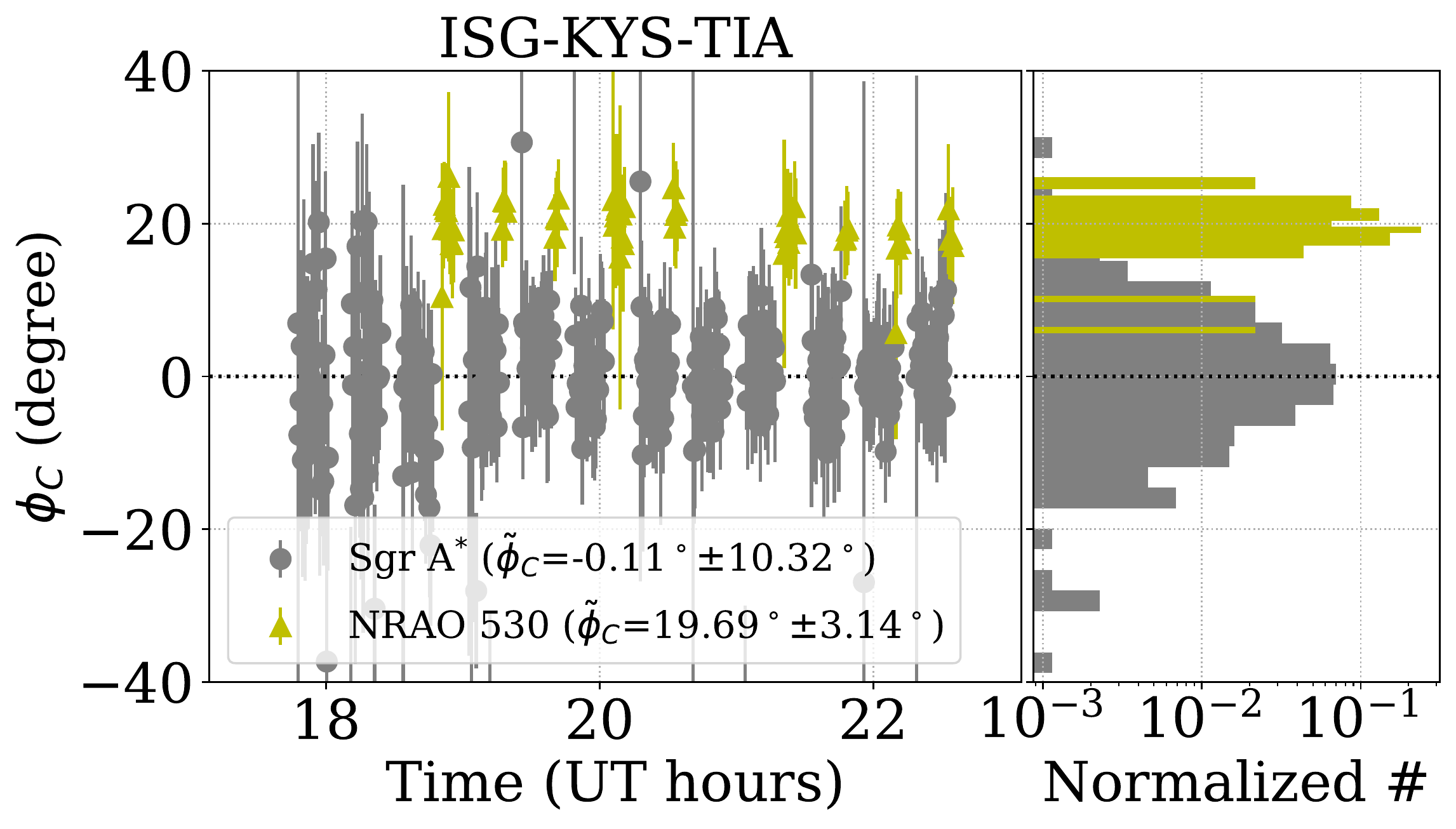} \\
\includegraphics[width=0.3\linewidth]{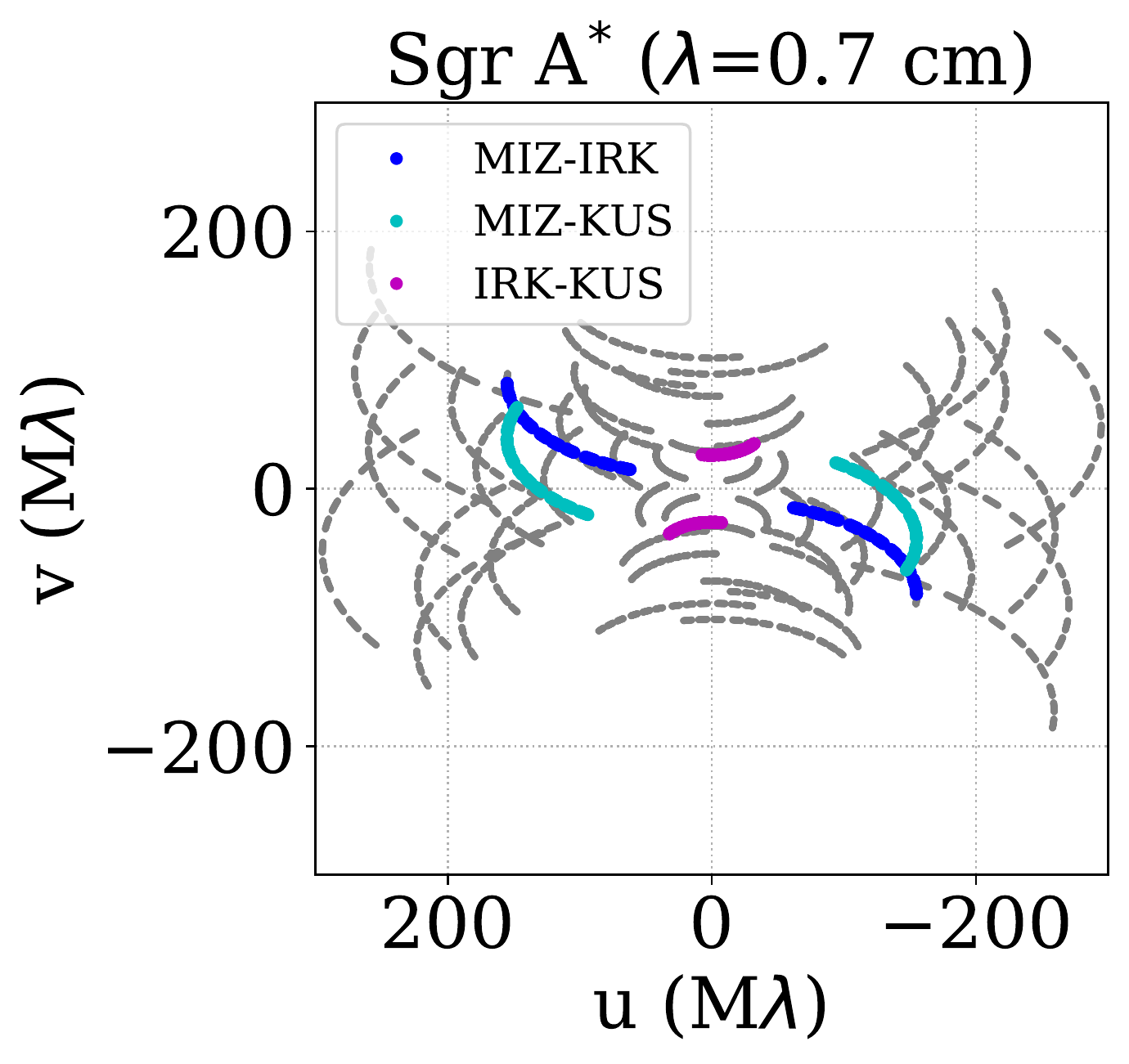} 
\includegraphics[width=0.5\linewidth]{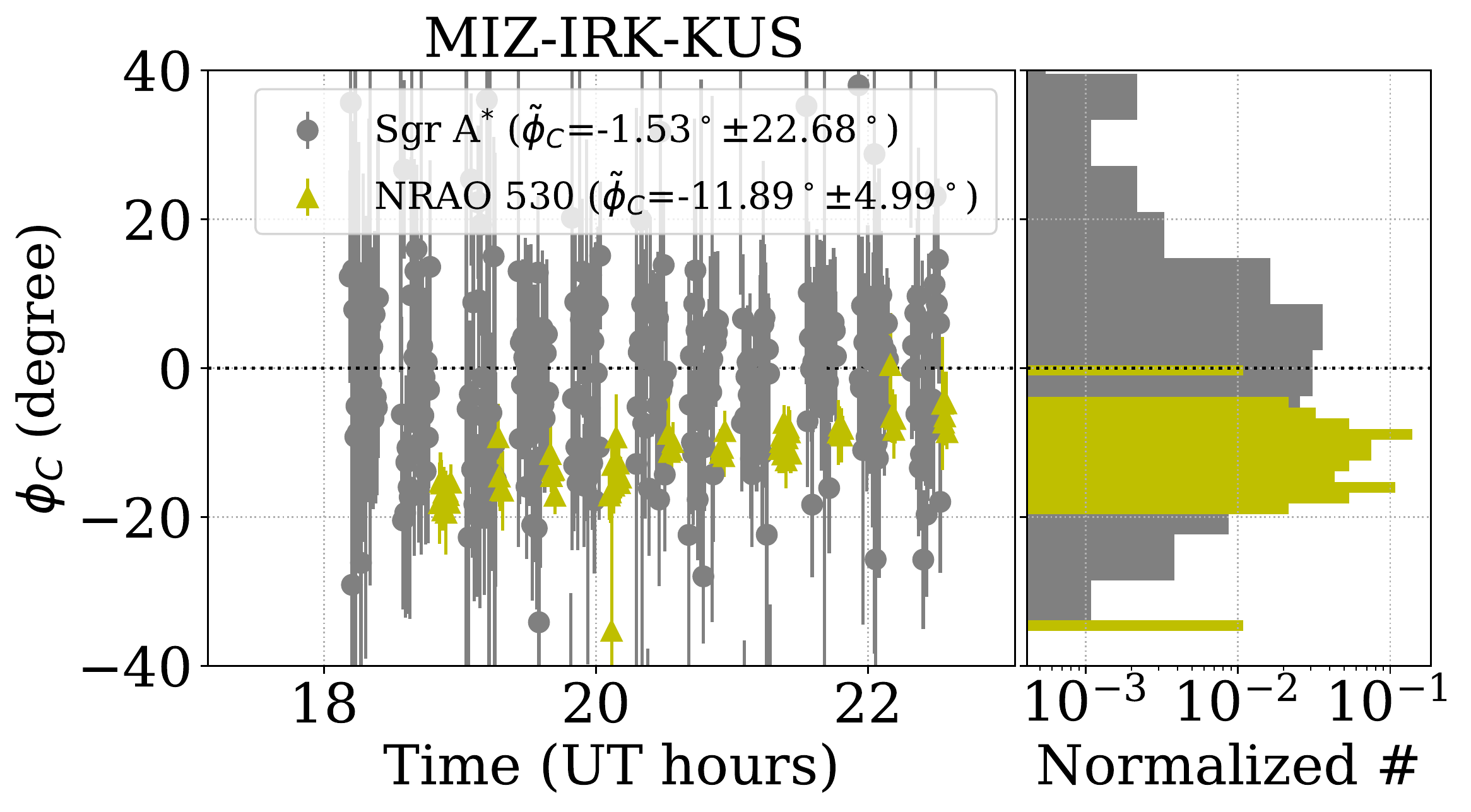} \\
\caption{
Closure phases of \sgra (gray) and NRAO\,530 (yellow) at 1.3\,cm (first and second rows) and 0.7\,cm (third and fourth rows). 
From left to right, the corresponding baselines towards \sgra in the $u-v$ plane (colored points), closure phases as a function of observing time, and histogram of the closure phases are shown. 
Each point has been averaged within 30\,seconds. 
The time averaged closure phase, $\tilde{\phi}_{c}$, is shown in each legend. The uncertainty represents the standard deviation.} 
\label{fig:cpplot}
\end{figure*}

The scattering effects toward \sgra can be approximated by a single thin phase screen \citep[e.g.,][]{Narayan_1989, Goodman_1989, Narayan_1992}. 
The interferometric visibility of scattering kernel is 
exp$\left[ -\frac{1}{2} D_{\phi} (b/(1+D/R)) \right]$, where $D_{\phi}$ is the phase structure function of the scattering screen, $b$ is the baseline length, $D$ is the distance between Earth and scattering screen, and $R$ is the distance between \sgra and scattering screen \citepalias{Johnson_2018}. 
For the baseline lengths $b \lesssim (1+D/R)r_{\rm in}$, where $r_{\rm in}$ is a finite inner scale of interstellar turbulence, the phase fluctuations are smooth so that the scatter broadening provides a Gaussian blurring (i.e., diffractive scattering dominated). 
At the baselines longer than this, on the other hand, both the scattering kernel and the refractive scattering noises introduce the non-Gaussian features (\citetalias{Psaltis_2018}, \citetalias{Johnson_2018}; \rev{see also, \autoref{sec:scat_modeling}}). 
Therefore we use the ``short'' baselines where can be fitted with a Gaussian model avoiding the noise biases. 
\rev{With $D=2.7\,$kpc, $R=5.4\,$kpc, $r_{\rm in}=800\,$km \citepalias{Johnson_2018}, the ``short'' baselines are found as $91$ and $176\,{\rm M}\lambda$ at 1.3\,cm and 7\,mm wavelengths, respectively.} 
Note however that the non-linear imaging reconstructions use the full baseline ranges (see \autoref{sec:selfcal}). 
In addition, the S/N cutoffs are also applied, for instance the S/N$_{\rm th}>3$ and S/N$_{\rm ref}>4$ where N$_{\rm th}$ and N$_{\rm ref}$ are thermal noise and refractive scattering noise, respectively (e.g., \citetalias{Johnson_2018}; see \autoref{sec:scat_modeling}, for the refractive noise derivation). 
This corresponds to the ensemble-average scattering limit (see \citealt{Goodman_1989, Narayan_1989}), where the scattering effects are \revnew{the} convolution of a scattering kernel onto an unscattered image. 

We add 10\,\% of visibility amplitudes in quadrature as systematic error for all KaVA data, which is typical amplitude gain uncertainty of cm-mm VLBI observations \citep[e.g.,][]{Cho_2017}. However, for TIA data at 1.3\,cm, we add 30\,\% of systematic errors to account for residual bandpass, relatively poorer system temperature measurements and larger pointing uncertainties (see, e.g., \citealt{Cui_2021} and EAVN status report\footnote{https://radio.kasi.re.kr/eavn/status\textunderscore report21/node3.html}). 

To confirm the single Gaussian approximation, we first check the closure phases which are the sum of visibility phases of three baselines forming a triangle. 
Since the closure phases are independent of the phase gain uncertainties, these are robust VLBI observables together with the closure amplitudes (e.g., \citealt{TMS, ehtc2019c, Blackburn_2020}; see also \autoref{sec:closure_amplitude}). 
We compare the closure phases of \sgra with a nearby calibrator, NRAO\,530. 
NRAO\,530 shows an extended structure towards \revnew{the} north and its closure phases are correspondingly deviated from zero, especially along \revnew{with} triangles that include long baselines (e.g., ISG-KYS).  
On the other hand, the closure phases of \sgra are consistent with zero (\autoref{fig:cpplot}) which indicates symmetric source structure. 
Therefore, we model the data with a single, elliptical Gaussian within \revnew{a} ``short'' baseline range (i.e., ensemble-average image). 
\\

\begin{figure}[t]
\centering 
\includegraphics[width=\columnwidth]{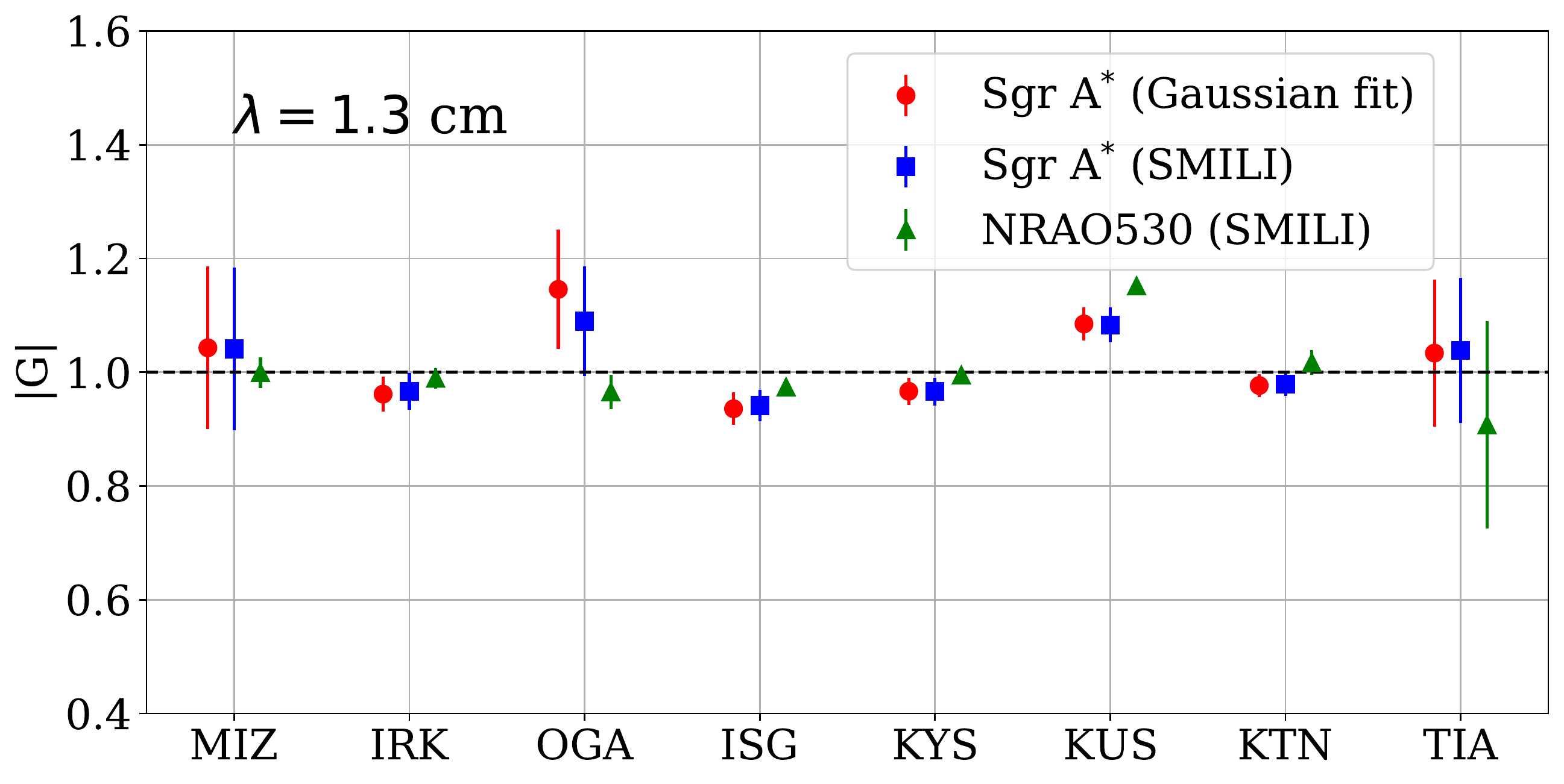} \\
\includegraphics[width=\columnwidth]{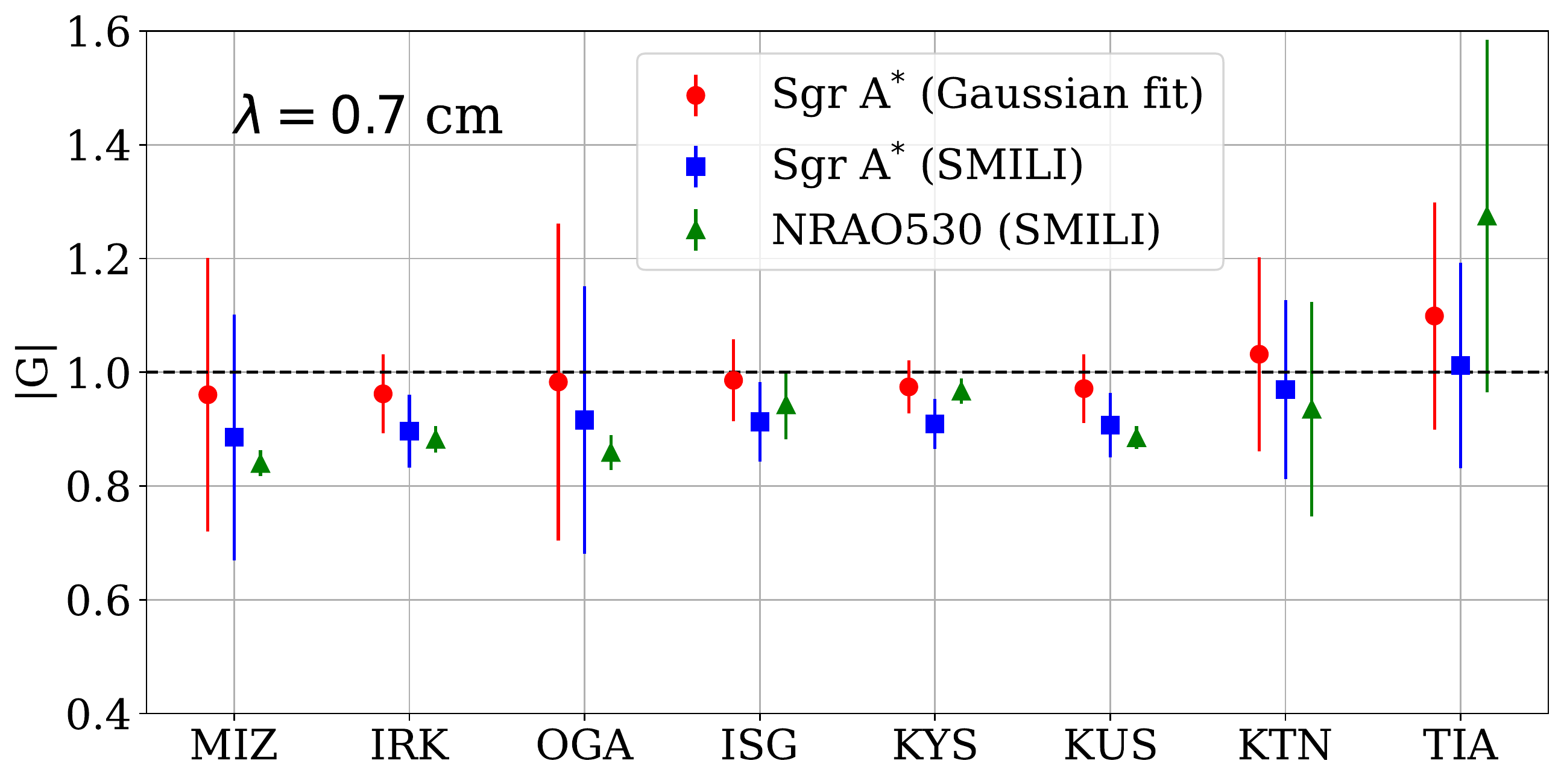}
\caption{
Multiplicative amplitude gain correction factors for each station at 1.3\,cm (upper) and 0.7\,cm (lower). The median values from Gaussian model fitting (red, circle) and SMILI imaging (blue, square) for \sgra are shown. The gain corrections for NRAO\,530 from SMILI imaging are also shown for comparison (green, triangle). 
The uncertainties are 1$\sigma$ standard deviation of the time dependent amplitude gain correction factors. 
}
\label{fig:gain_correction}
\end{figure}

\begin{figure*}[t]
\centering 
\includegraphics[width=0.49\textwidth]{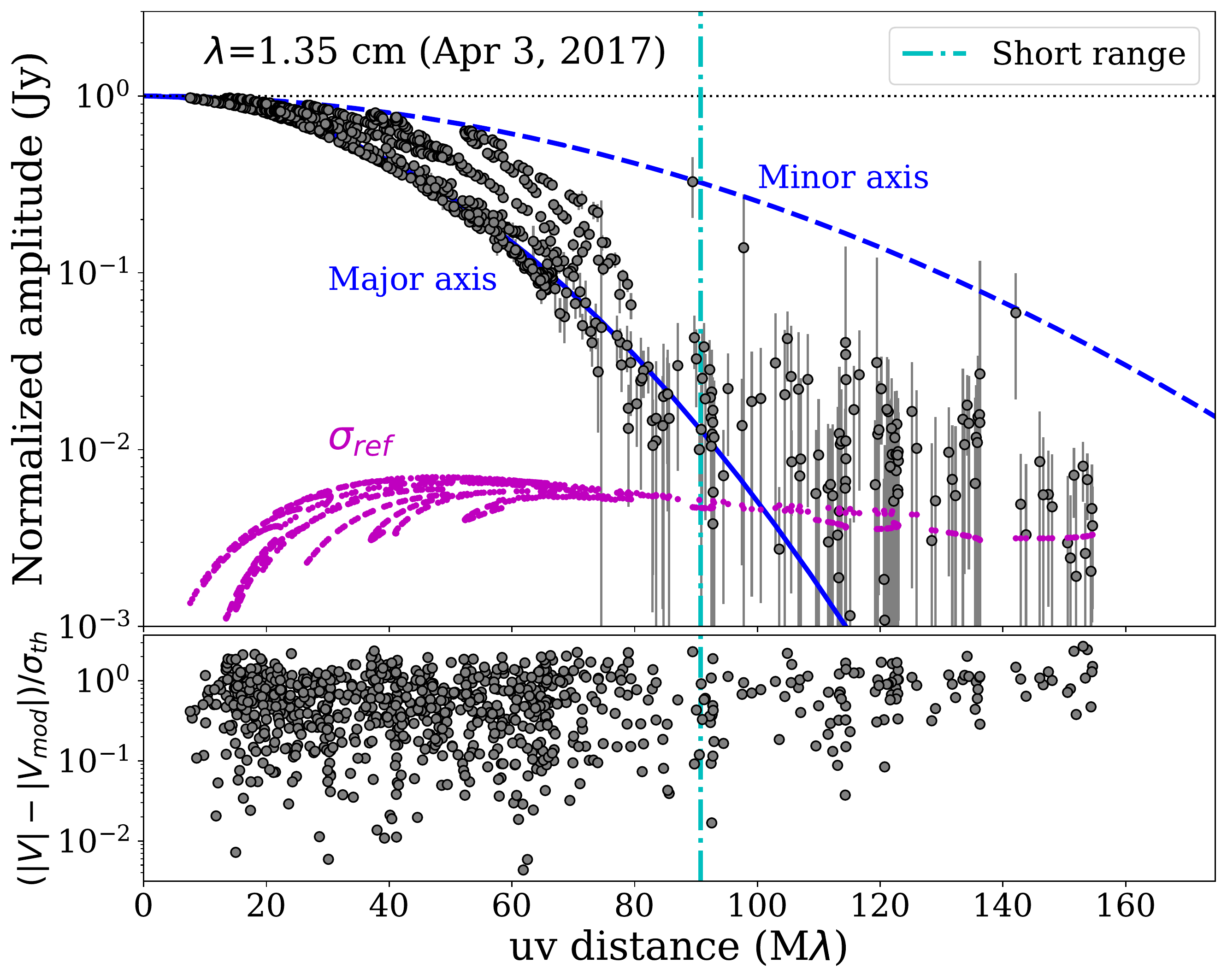}
\includegraphics[width=0.49\textwidth]{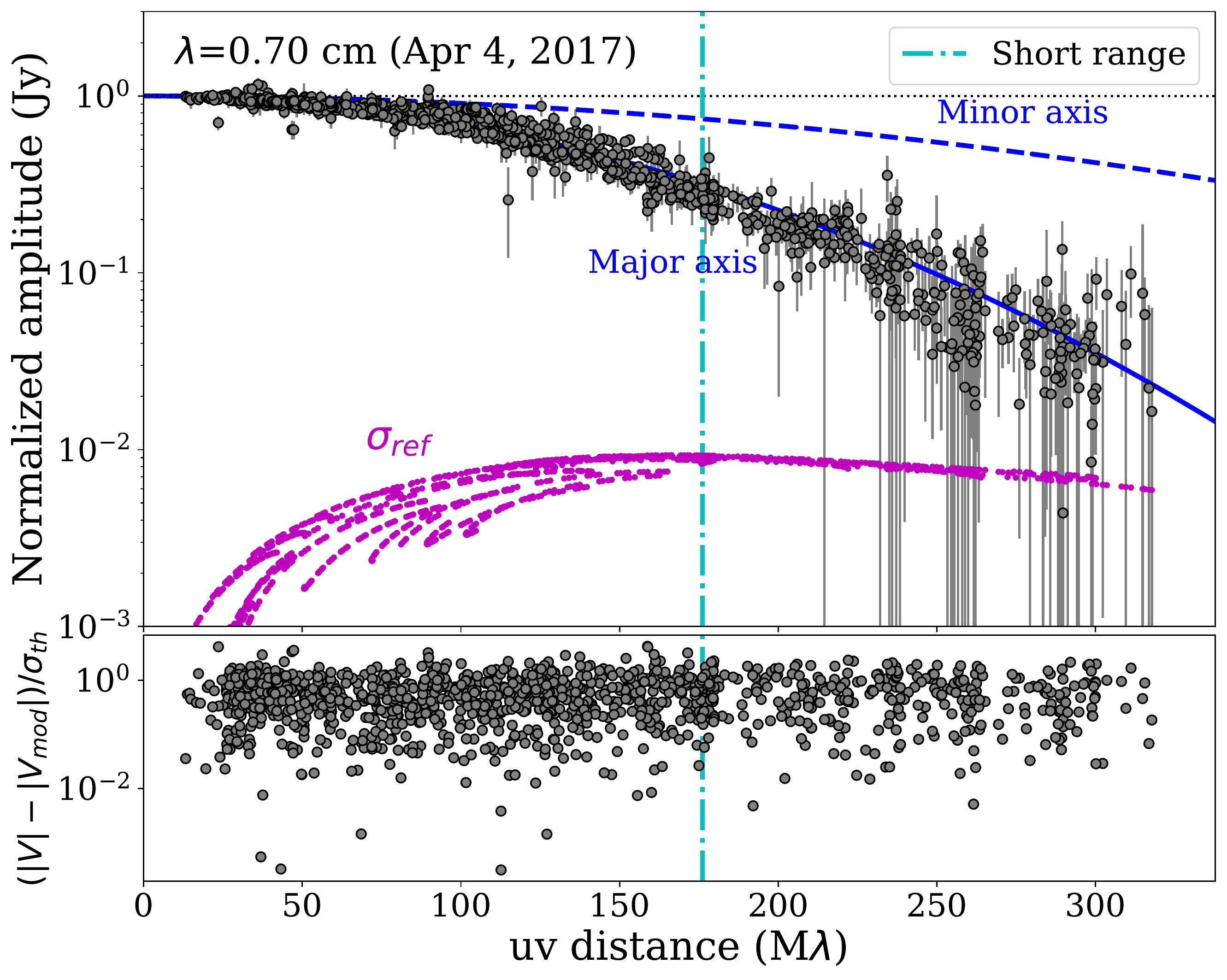}
\caption{
(Upper) Normalized correlated flux density of \sgra at 1.3\,cm (left) and 0.7\,cm (right). 
The gray circles are visibility amplitudes after self-calibration. 
The blue lines are the Gaussian scattering kernel towards \revnew{the} major axis (solid line) and minor axis (dashed line). 
The cyan-colored, broken-dotted vertical line shows the ``short'' range at each wavelength. 
The magenta points are the derived refractive noise, $\sigma_{\rm ref}$. 
(Lower) The ratio of the residual of the visibility amplitudes and the fitted model amplitudes, over the thermal noise. Each point has been 5\,minutes averaged for clearer visualization.} 
\label{fig:selfcal_radpl}
\end{figure*}

\begin{figure*}[t]
\centering 
\includegraphics[width=\columnwidth]{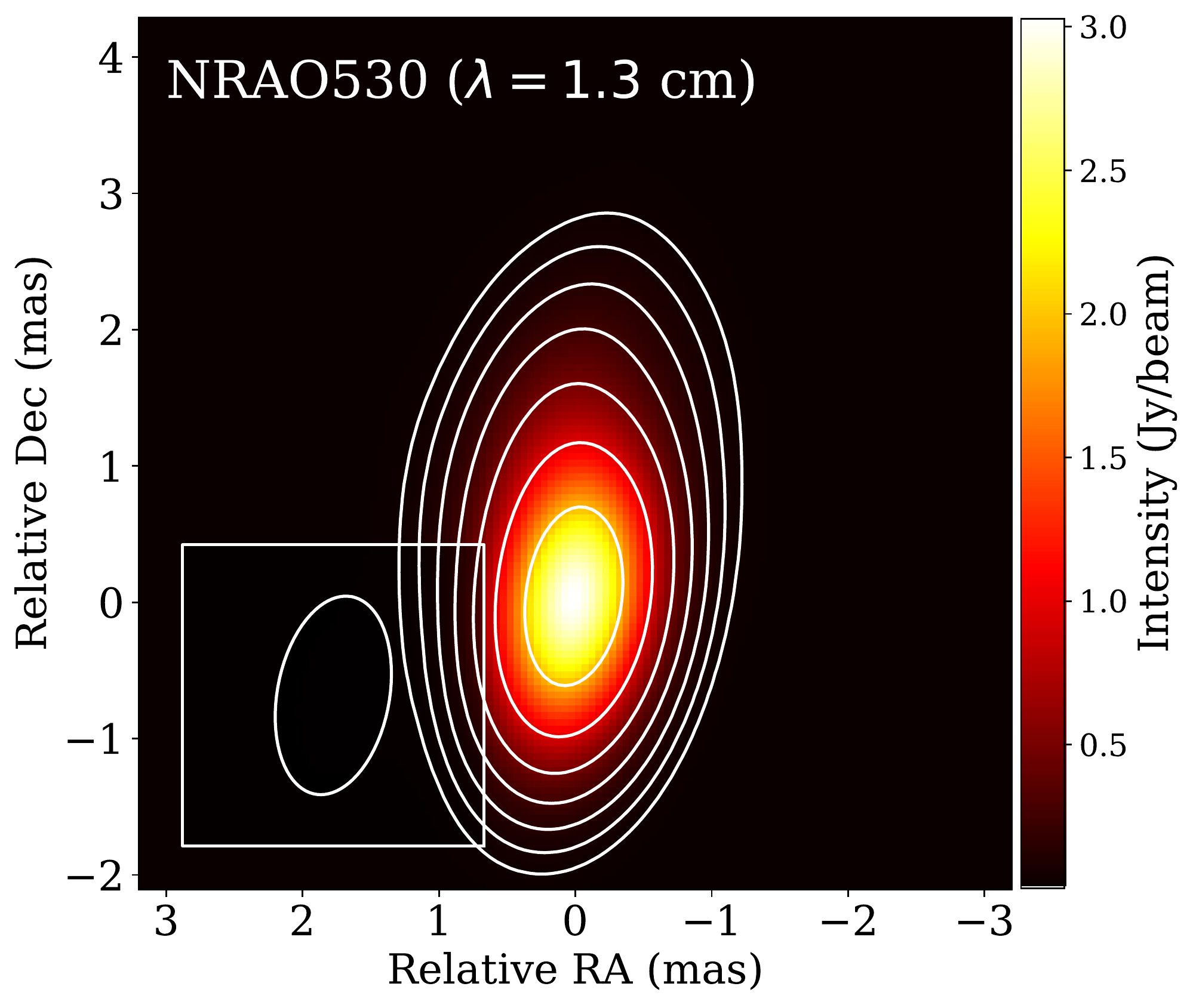}
\includegraphics[width=\columnwidth]{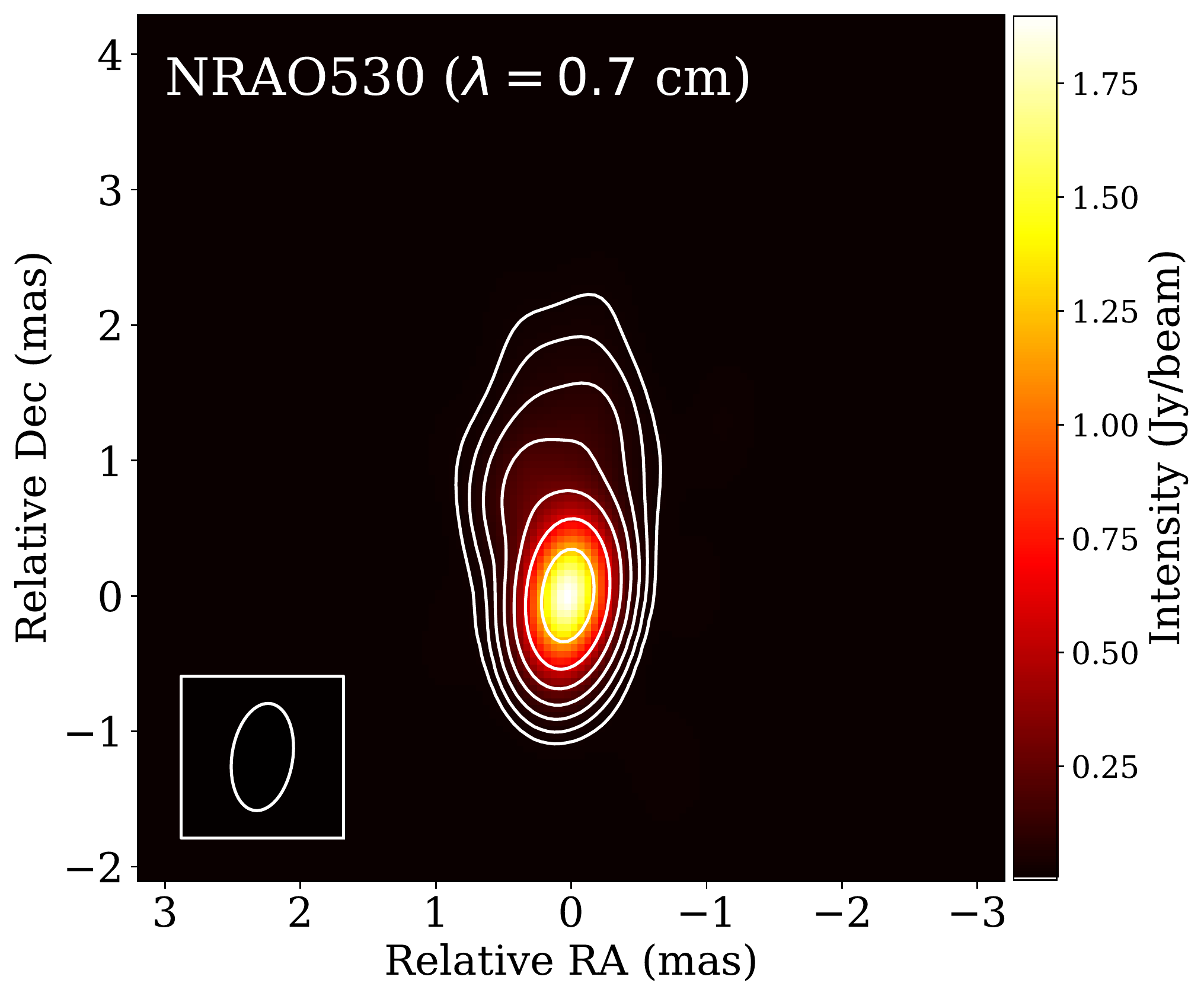} \\
\includegraphics[width=\columnwidth]{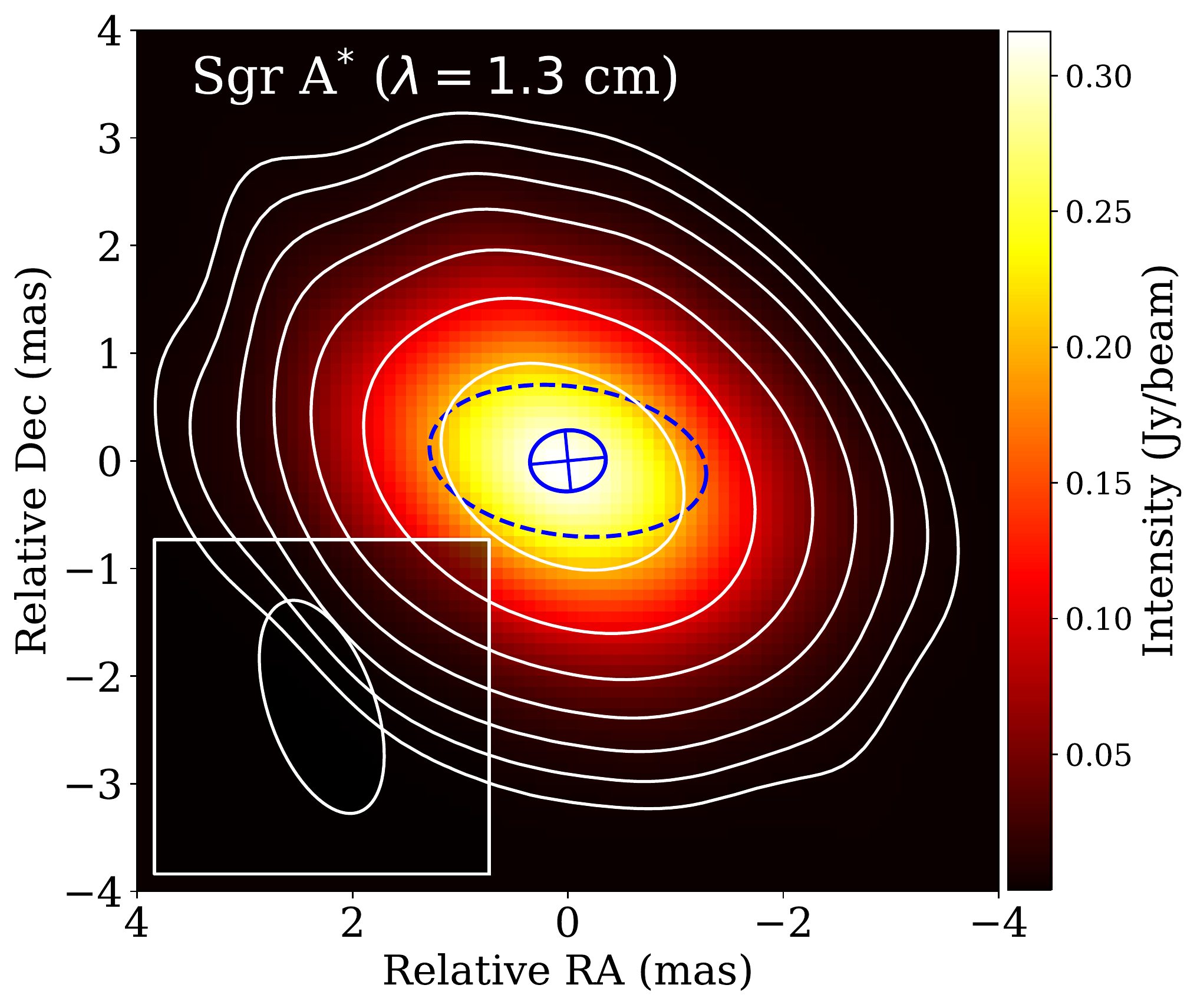}
\includegraphics[width=\columnwidth]{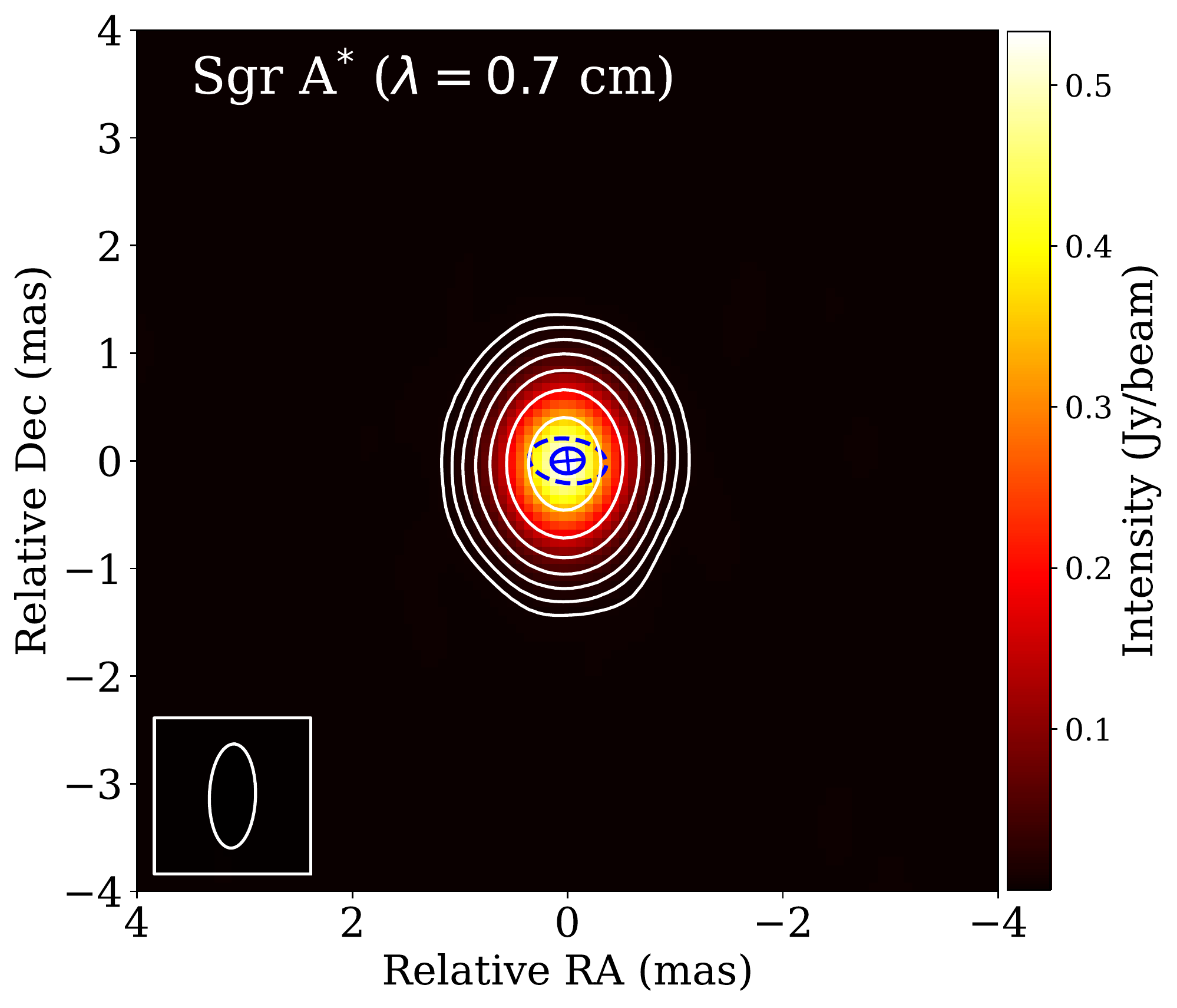}
\caption{
Observed structure of NRAO\,530 (upper) and \sgra (lower), at 1.3\,cm (left) and 0.7\,cm (right). The restoring beam is shown in the lower-left side of each panel. The contour levels are (1, 2, 4, 8, 16, 32, 64)\,\% of peak intensity. 
The blue ellipses \rev{with broken-line and solid-line (with cross)} at the center of \sgra images show the best-fitted Gaussian model of  \rev{ensemble-average and intrinsic} structure, \rev{respectively, derived from closure amplitudes (see \autoref{tab:results})}. 
As for the NRAO\,530, the extended jet feature is well consistent with the one seen in previous studies \citep[e.g.,][]{Lu_2011b, Jorstad_2017}. 
}
\label{fig:imdifmap}
\end{figure*}

\subsection{Self-calibration}
\label{sec:selfcal}

For the self-calibration which corrects the station-based gain uncertainties, we reconstruct the image of \sgra using two different approaches: 
1) Gaussian model fitting to the complex visibilities using the {\tt modelfit} in DIFMAP software \citep{Shepherd_1994}, and 
2) imaging based on both the complex visibilities and closure quantities using the regularized maximum likelihood (RML) method. 

As for the two-dimensional (2D) Gaussian model fitting, a single, elliptical Gaussian model is fitted within the ``short'' baseline range, $\lesssim\,91$ and $\lesssim\,176\,{\rm M}\lambda$ at 1.3\,cm and 7\,mm wavelengths, respectively. 
Note that these are $\sim$ half of the longest baseline lengths where the fringes toward \sgra have been successfully detected. 
Nevertheless, we can still derive gain corrections for all stations using only the data in this range since every station has at least one independent closure amplitude on most of the scans. 

For the imaging via RML method, we use SMILI software library\footnote{https://github.com/astrosmili/smili} \citep{Akiyama_2017a, Akiyama_2017b, Smili_2019}. 
\rev{
SMILI has more flexibility in the data products used for imaging, so} 
this enables us to use the robust closure quantities. 
To reconstruct the observed structure of \sgra, therefore, both the visibility amplitudes and closure quantities are used. 
For SMILI imaging, \revnew{a} full baseline range has been used so that the refractive sub-structures are also reconstructed and the visibilities are self-calibrated with this model. Note that the refractive noises are baseline-based so that they have a minor effect on the derived station-based gain uncertainties. 
The imaging parameters have been searched for the regularizers of $\ell_1$ norm and total squared variation (TSV), \revnew{which represent the image sparsity and smoothness, respectively,} in a range of [0.01, 0.1, 1.0, 10, 100] 
\rev{(see \autoref{app:smili_imaging}, for SMILI imaging and its regularizers).} 
The \rev{fixed values of} $\ell_1$ prior = 2 and field-of-view = 12.8\,mas (128\,pixels$\,\times\,$0.1\,mas/pixel) are used. 
Based on the quadratic sum of $\chi^{2}$ of closure quantities, the fiducial parameters for \sgra are found as [$\ell_1$-norm, TSV] = [0.1, 0.1] and [0.01, 0.1] at 1.3 and 0.7\,cm, respectively. 
For NRAO\,530, [$\ell_1$-norm, TSV] = \rev{[0.01, 0.1] and [0.1, 1.0] at 1.3 and 0.7\,cm, respectively.}

As a result, the derived multiplicative gains for each telescope are well consistent between two different methods (\autoref{fig:gain_correction}). 
For comparison, the gain solutions from NRAO\,530 are also derived by the SMILI imaging. 
\autoref{fig:selfcal_radpl} shows the self-calibrated visibility amplitudes with the Gaussian model fitting. Note that the non-Gaussian noises appear at the longer baselines.   
The sizes of ensemble-average image are then obtained by Gaussian model fitting onto the self-calibrated visibility amplitudes within the ``short'' range: $\sim2.6\times1.4$ mas (position angle $\sim\,82^\circ-83^\circ$) and $\sim0.7\times0.4$ mas  (position angle $\sim\,83^\circ-85^\circ$) at 1.3\,cm and 7\,mm, respectively (see \autoref{tab:results}). 
These are consistent with the scattering dominated size of \sgra, with the asymptotic Gaussian scattering kernel of $(1.380\pm0.013)\lambda^2$ (mas) for major-axis and $(0.703\pm0.013)\lambda^2$ (mas) for minor-axis where $\lambda$ is observing wavelength in cm \citepalias{Johnson_2018}. 
For instance, the asymptotic Gaussian scattering kernel sizes at 1.3 and 0.7\,cm are $\sim2.51\times1.28$\,mas and $\sim0.67\times0.34$\,mas respectively, which are slightly smaller than the measured ensemble average sizes. 
\autoref{fig:imdifmap} shows the reconstructed images of NRAO\,530 and \sgra from SMILI imaging. Note that the \sgra images show scattered, not intrinsic, structure. The contour and color maps show the beam convolved structure. 
\rev{The ellipses at the center of \sgra image show the best-fitted 2D Gaussian model of ensemble-average (broken-line) and intrinsic (solid-line, cross) structure of \sgra, from the closure amplitudes (see \autoref{subsec:intrinsic_size} and \autoref{tab:results}).} 

\subsection{(Log) Closure amplitudes}
\label{sec:closure_amplitude}
While the best-fitted model to visibility amplitudes can provide reasonable antenna gain solutions, it may be still difficult to take into account all possible gain uncertainties \citep{Bower_2014}. To overcome this, we fit the 2D Gaussian model to the log closure amplitudes which are robust observables free from the amplitude gain uncertainties. 
Note however that it is difficult to constrain the flux density with only the closure amplitudes so it is complementary to the self-calibration and vice-versa. 

The closure amplitudes of four different stations ($i,j,k,l$) are defined as the ratios of visibility amplitudes so that the station-based amplitude gain errors are \revnew{canceled} out. Three different closure amplitudes can be formed with any set of four stations, and two of them are independent (e.g., \citealt{Chael_2018}), 

\begin{equation}
    C_{1} = \frac{|V_{ij}||V_{kl}|}{|V_{ik}||V_{jl}|}, 
    C_{2} = \frac{|V_{ik}||V_{jl}|}{|V_{il}||V_{jk}|}, 
    C_{3} = \frac{|V_{il}||V_{jk}|}{|V_{ij}||V_{kl}|}, 
\label{eq:camp}
\end{equation}
where the $C$ is the closure amplitude at each quadrangle and $|V|$ is the visibility amplitude at each baseline. 
The number of independent closure amplitudes at a certain time range is $N(N-3)/2$, where the $N$ is the number of stations. 
There are a number of ways to form the independent closure amplitudes, and it is important to select the set where the S/N is high enough to limit the correlated noise biases \citep{Blackburn_2020}. 

Using SMILI, we form the independent closure amplitudes from the visibilities within the aforementioned ``short'' range. The closure amplitudes of 2D Gaussian model are formed in the same way. The best fitted model is found using a Monte Carlo method
\footnote{A Gaussian model is provided with the free parameters of major axis size, axial ratio, and position angle. Its amplitude is then given at each $u$, $v$ point which corresponds to the observed data. The random sampling of the free parameters are repeated thousands times, for instance using the No-U-Turn Sampler (NUTS) which is a particular Markov Chain Monte Carlo (MCMC) algorithm, and finds the optimized numerical solution for the searching parameters.} 
onto the log closure amplitudes to avoid the biases and to symmetrize the numerator and denominator in closure amplitudes \citep[e.g.,][]{Chael_2018, Blackburn_2020}. 
Note that the closure amplitudes are insensitive to the total flux density so it is not fitted but rather given as the value obtained from the self-calibrations. 
Therefore three parameters of ensemble-average image, major axis size ($\theta_{\rm maj}^{\rm en}$), axial ratio ($r^{\rm en}\equiv\theta_{\rm maj}^{\rm en}$/$\theta_{\rm min}^{\rm en}$), and position angle ($\theta_{\rm PA}^{\rm en}$) have been estimated. 
From this, we find that the results are consistent with the amplitude self-calibration (\autoref{tab:results}). 
This demonstrates that the reconstructed models from imaging and 2D Gaussian model fitting properly corrects the station-based gain uncertainties. 
\\

\subsection{Scattering kernel model and deblurring} 
\label{sec:scat_modeling}

According to the physically motivated scattering model, the scattering kernel is no longer a simple (anisotropic) Gaussian and not purely proportional to the square of observing wavelengths \citepalias{Psaltis_2018}. 
Adopting the \citetalias{Psaltis_2018} model, therefore, we have developed the {\tt scatter model} module of the SMILI\footnote{\rev{https://github.com/astrosmili/smili/tree/v0.1/smili/scattering}}. 
This is almost identical with the {\tt stochastic optics} module of the {\tt eht-imaging} software library\footnote{https://github.com/achael/eht-imaging} \citep{Johnson_2016,Chael_2018}, and the scattering parameters have been adopted from \citetalias{Johnson_2018} \rev{(e.g., $\alpha=1.38^{+0.08}_{-0.04}$, $r_{\rm in}=800\pm200\,$km)}. 
Then the self-calibrated, ensemble-average visibility amplitudes and closure amplitudes are divided by the derived scattering kernel (i.e., deblurring) which corresponds to the deconvolution in the image domain. 

To estimate the refractive noises, we first generate the synthetic images (e.g., intrinsic structure of \sgra from \citetalias{Johnson_2018}) and scatter \revnew{them} $\sim$1,000 times each with a randomly generated phase screen using the {\tt stochastic optics} module. 
The corresponding visibilities are then obtained by Fourier transform at each $u, v$ point of the EAVN data. As the number of iteration increases, the standard deviation of real and imaginary parts of the visibilities converge to the root-mean-squared (rms) refractive noise at each $u, v$ point. 
The derived refractive noises are used to constrain the noise-dominated visibilities, through the aforementioned S/N$_{\rm ref}$ cutoffs. \\

\section{Results} \label{sec:results}

\begin{table*}[ht]
\centering
\addtolength{\tabcolsep}{-3pt}
\small
\caption{Gaussian model fitting results}
\begin{tabular}{lcc cccccccc}
\hline
\hline
$\lambda$ (cm) & 
Method & 
S$_{\nu}^{\rm tot}$ (Jy) & 
$\theta_{\rm maj}^{\rm en}$ ($\mu$as) & 
$\theta_{\rm min}^{\rm en}$ ($\mu$as) & 
${r}^{\rm en}$ & 
$\theta_{\rm PA}^{\rm en}$ (deg) &
$\theta_{\rm maj}^{\rm int}$ ($\mu$as) & 
$\theta_{\rm min}^{\rm int}$ ($\mu$as) & 
${r}^{\rm int}$ & 
$\theta_{\rm PA}^{\rm int}$ (deg) 
\\
\hline
1.349 & Gfit/Amp & 1.05$\,\pm\,$0.11 & 
2621.1$\,\pm\,$26.4 &  1424.4$\,\pm\,$13.3 &  1.84$\,\pm\,$0.02 &  82.8$\,\pm\,$ 0.4 & 
827.1$\,\pm\,$93.4 &  630.8$\,\pm\,$52.6 &  1.31$\,\pm\,$0.19 &  95.6$\,\pm\,$23.4   \\
 & SMILI/Amp & 1.04$\,\pm\,$0.10 & 
2617.9$\,\pm\,$26.4 &  1412.7$\,\pm\,$13.5 &  1.85$\,\pm\,$0.02 &  82.8$\,\pm\,$ 0.4 & 
814.8$\,\pm\,$93.3 &  602.6$\,\pm\,$53.1 &  1.35$\,\pm\,$0.19 &  94.6$\,\pm\,$23.4  \\
 & CA & ... & 
2585.1$\,\pm\,$27.9 &  1383.3$\,\pm\,$30.2 &  1.87$\,\pm\,$0.05 &  82.5$\,\pm\,$ 0.6 & 
704.3$\,\pm\,$102.0 &  566.7$^{+78.1}_{-84.8}$ &  1.19$^{+0.24}_{-0.19}$ &  95.5$\,\pm\,$27.8   \\
\hline
0.695 & Gfit/Amp & 1.36$\,\pm\,$0.14 & 
715.6$\,\pm\,$8.6 &  414.9$\,\pm\,$8.5 &  1.73$\,\pm\,$0.04 &  84.2$\,\pm\,$0.7 & 
294.6$\,\pm\,$23.6 &  229.3$^{+12.8}_{-13.7}$ &  1.28$\,\pm\,$0.11 &  109.6$\,\pm\,$8.9   \\
 & SMILI/Amp & 1.30$\,\pm\,$0.13 & 
727.2$\,\pm\,$8.6 &  425.7$\,\pm\,$8.2 &  1.71$\,\pm\,$0.04 &  85.6$\,\pm\,$0.7 & 
331.1$\,\pm\,$23.4 &  235.5$\,\pm\,$12.0 &  1.40$\,\pm\,$0.11 &  112.7$\,\pm\,$7.3   \\
 & CA & ... & 
720.8$\,\pm\,$9.0 &  412.0$\,\pm\,$19.8 &  1.75$\,\pm\,$0.09 &  83.2$\,\pm\,$1.0 & 
300.0$\,\pm\,$24.8 &  231.0$^{+32.3}_{-34.7}$ &  1.28$\,\pm\,$0.20 &  95.2$^{+13.6}_{-12.3}$  \\
\hline
\end{tabular}
\\ \vspace{0.3cm}
\raggedright{\textbf{Note. }
Overall results of total flux density, scattered and unscattered size of \sgra at 1.3 and 0.7\,cm wavelengths. 
From left to right: observing wavelength, imaging/model-fitting method, total flux density, ensemble-average (i.e., scattered) structure and intrinsic (i.e., unscattered) structure. 
The results are from three different methods: 
self-calibrate with the Gaussian model fitting and size fitting on the visibility amplitudes (Gfit/Amp), 
self-calibrate with the SMILI imaging and size fitting on the visibility amplitudes (SMILI/Amp), and 
the size fitting on the log closure amplitudes (CA). See \autoref{sec:imfit} for more detail. 
The subscripts are specific frequency ($\nu$), major axis (maj), minor axis (min), and position angle (PA). 
The superscripts are total flux density (tot), ensemble-average image properties (en) and intrinsic properties (int). 
The $\pm1\,\sigma$ uncertainties are shown considering the possible error components (see \autoref{app:error}), except the total flux density which shows the 10\,\% of each measurement.} 
\label{tab:results}
\end{table*}

Here we summarize the observational results. 
The obtained intrinsic sizes of \sgra and its wavelength-dependence are presented in \autoref{subsec:intrinsic_size}. 
In \autoref{subsec:flux_densities}, the flux densities and spectral indices of \sgra are shown.

\subsection{$\lambda-$dependent intrinsic size of \sgra} \label{subsec:intrinsic_size}

The intrinsic structure of \sgra is found from the Gaussian model fitting onto the deblurred visibility amplitudes or closure amplitudes. While the ``observed'' size is angular broadened by the diffractive scattering, the ``intrinsic'' size is scatter-deblurred so that it is free from the angular broadening. 
The Monte Carlo method is used to find the optimal values of three free parameters: ``intrinsic'' major axis size ($\theta_{\rm maj}^{\rm int}$), axial ratio ($r^{\rm int}\equiv\theta_{\rm maj}^{\rm int}$/$\theta_{\rm min}^{\rm int}$), and position angle ($\theta_{\rm PA}^{\rm int}$). 
Note that the scattering deblurred amplitudes are normalized with the total flux density, S$_{\nu}^{\rm tot}$. 
The axial ratio of major and minor axis size is $\sim1.2-1.4$ and the PA is $\sim95-113^\circ$, so the intrinsic sizes of \sgra at both wavelengths look slightly elongated towards \revnew{the} east-west direction. 
The PA is consistent with the previous studies (e.g., $\sim105^{\circ}$ from \citealt{Markoff_2007}; $\sim95^{\circ}$ from \citealt{Bower_2014}), and roughly aligned with the orientation of the large-scale X-ray emission as well \citep{Li_2013}. 
Considering the uncertainties, however, the axial ratios are consistent with unity within the $\lesssim3\sigma$ uncertainty range (\autoref{tab:results}). 
See \autoref{app:error} for more details of the uncertainty estimates. 
While this is consistent with the \citetalias{Johnson_2018}, \citet{Bower_2014} reported the axial ratio $\sim2.8$ coming from the smaller minor axis size at 7\,mm, $\sim126\,\mu$as. This may indicate the possible time variation of the minor axis size of \sgra, and it will be investigated through the long-term monitoring observations (e.g., X.~Cheng et al. in prep.).  
Note that the results from different methods are well consistent with each other within the uncertainties. 
For later discussions, the results from model fitting onto the log closure amplitudes are used since this provides more robust measurement and conservative uncertainties. 

\begin{figure}[t]
\centering 
\includegraphics[width=\columnwidth]{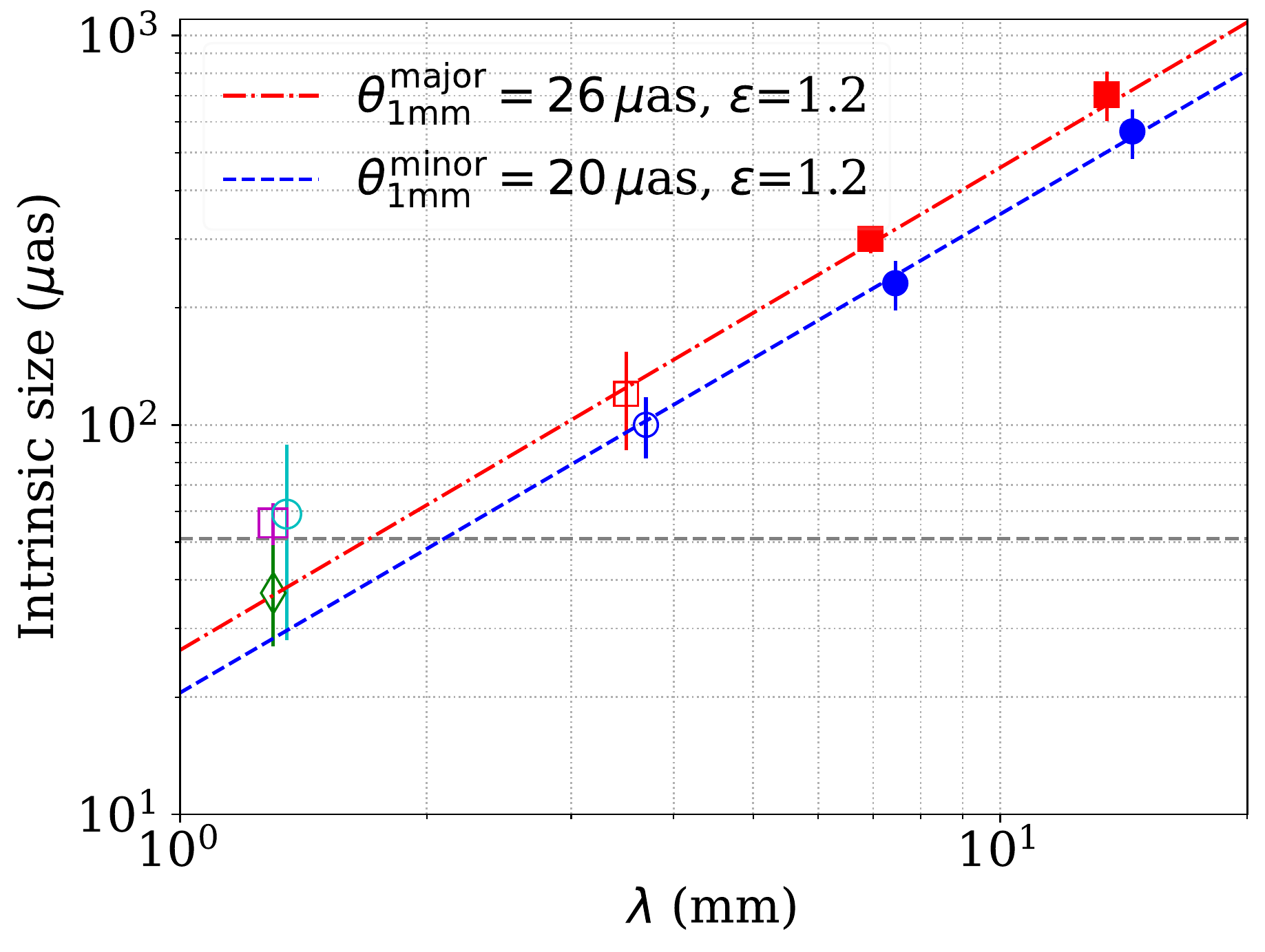}
\caption{
Intrinsic size of \sgra as a function of wavelength. 
The squares and circles show the major and minor axis sizes (i.e., full-width half-maximum), respectively. 
While the filled markers are the results from this work, the open markers show the previous experiments at 3.5\,mm \citepalias{Issaoun_2019} and 1.3\,mm (\citetalias{Johnson_2018}; \citealt{Doeleman_2008} with green, diamond). 
The markers are offset slightly in wavelength for clearer visualization. 
The power law fitting results are shown: towards the major axis (red, broken-dotted line) and minor axis (blue, broken line) sizes. 
Note that only the 3.5, 7, and 13\,mm wavelengths are used for the fitting which are the quasi-simultaneous measurements in 2017 April. 
The 1.3\,mm results are shown for comparison (e.g., the extrapolated sizes). 
The horizontal broken line (gray-colored) shows the expected black hole shadow diameter at \sgra ($\sim51\,\mu$as; e.g., \citetalias{Johnson_2018}). 
}
\label{fig:results}
\end{figure}

Together with the intrinsic size of \sgra at 3.5\,mm which has been measured in $\sim$ a day separation from our EAVN observations \citepalias{Issaoun_2019}, we fit the wavelength-dependent intrinsic size with a power law, $\theta_{\lambda} = \theta_{\rm 1mm}\times\lambda_{\rm mm}^{\epsilon}\,(\mu{\rm as})$ where the $\theta_{\rm 1mm}$ is the source size at 1\,mm, $\lambda_{\rm mm}$ is the observing wavelength in mm, and $\epsilon$ is the power-law index which characterizes the intrinsic properties of the emission region as a function of observing wavelength. 
As a result, the measured sizes of \sgra at 13, 7, 3.5\,mm are well fitted with a single power law (see \autoref{fig:results}). 
The derived power law index, $\epsilon=1.2\pm0.2$, is consistent with the previous experiments of $\epsilon\sim1.2-1.3$ (e.g., \citealt{Falcke_2009}; \citealt{Lu_2011}; \citealt{Ortiz-Leon_2016}; \citealt{Brinkerink_2019}) and $\epsilon\sim1.0-1.1$ (e.g., \citealt{Shen_2005}, \citetalias{Johnson_2018}). 
Note that the derived $\epsilon$ from \revnew{size fitting results on the visibility amplitudes, which are self-calibrated with the Gaussian model fitting (Gfit/Amp) and SMILI imaging (SMILI/Amp),} are $1.4\pm0.2$ and $1.3\pm0.2$ respectively. 
\revnew{These} are consistent with the result from \revnew{closure amplitudes (CA)} within each uncertainty. 
The uncertainties of our fitted results are relatively large since we only use the (quasi-) simultaneous observations so that the number of data points is small. 
To better constrain the size-wavelength relation, therefore, the intrinsic sizes at more various wavelengths are necessary especially at 2\,mm and $\gtrsim2\,$cm. 
In addition, the quasi-simultaneous campaigns at a variety of wavelengths, as we have achieved in 2017 together with the GMVA+ALMA and EHT, are further required since the (marginal) size variation of \sgra, especially at 7\,mm, has been found \citep[e.g.,][]{Bower_2004, Akiyama_2013, Zhao_2017}. 

From the best-fitted results, the extrapolated size at 1.3\,mm is found as \rev{$36^{+18}_{-16}\,\mu$as and $27^{+13}_{-12}\,\mu$as} toward major and minor axis, respectively. 
This is close to the previous measurements with EHT (e.g., 
$37\pm13\,\mu$as from \citealt{Doeleman_2008}; 
$41-44\,\mu$as from \citealt{Fish_2011}; 
$51-52\,\mu$as from \citealt{Lu_2018}; 
$56\pm7\,\mu$as and $59^{+30}_{-31}\mu$as for major and minor axis, respectively, from \citetalias{Johnson_2018}). 
Note however that the structure of \sgra at 1.3\,mm may be non-Gaussian \citep[e.g.,][]{Johnson_2015,Fish_2016, Lu_2018} so that the extrapolated size can be a hint for high resolution imaging, for instance with the EHT. 
\\

\subsection{Flux densities and spectral indices}
\label{subsec:flux_densities}

At radio to sub-mm wavelengths, \sgra shows an inverted (time-averaged) spectrum, $S_{\nu}\propto\nu^{\beta}$ ($\beta>0$; e.g., \citealt{Duschl_Lesch_1994, Morris_1996, Serabyn_1997, Falcke_1998, Krichbaum_1998, Zhao_2001}) where $S_{\nu}$ is specific flux density, $\nu$ is the observing frequency, and $\beta$ is the spectral index. 
The spectrum peaks and cutoffs are shown at $230-350\,$GHz \citep[e.g.,][]{Yusef-zadeh_2006b, Bower_2015}, which are due to the transition of synchrotron emission from being optically thick to thin. 
This implies that \sgra has a stratified, self-absorbed geometry of plasma, either a jet or an accretion flow, and the (sub-)\,mm emission \revnew{arise} from several $R_{\rm s}$ \citep[e.g.,][]{Melia_1992, Melia_1994}. 
At mm/sub-mm wavelengths, in addition, there exists a break in the spectrum so called ``mm/sub-mm bump" which deviates from a single power-law index. This may be explained by the compact components in the acceleration zone of a jet, or the thermal electrons in an accretion flow \citep[e.g.,][]{Lu_2011}.

In our observations, the total flux densities are $\sim$1.0 and 1.3\,Jy at 22 and 43\,GHz, respectively (see \autoref{tab:results}). 
With the intrinsic major axis size, we can derive a lower limit of the brightness temperature as $T_{\rm b} \gtrsim 0.4-0.5\times10^{10}\,$K and $\gtrsim 0.8-1.0\times10^{10}\,$K at 1.3 and 0.7\,cm respectively. 
Together with the flux density at 86\,GHz, $\sim2\,$Jy \citepalias{Issaoun_2019}, the spectral index of \sgra within the three frequencies is derived as $\beta=0.39\pm0.02$. 
Considering the mm/sub-mm bump at $\gtrsim43\,$GHz, at the same time, we derive the values of $\beta$ for 22$-$43\,GHz and 43$-$86\,GHz separately and find $\beta\approx\,$0.29$\pm$0.10 and 0.44$\pm$0.16 respectively. 
Note that the smaller number of points \revnew{leads} to larger uncertainties. 
Nevertheless, the results are well consistent with the historical studies. 
For instance, while $\beta\sim0.24-0.3$ up to $\sim43\,$GHz \citep[e.g.,][]{Krichbaum_1998, Falcke_1998, Lu_2011, Bower_2015}, $\beta\sim0.5$ at $\gtrsim43\,$GHz \citep{Falcke_1998, Bower_2015}. 

With the derived spectral indices, the flux density at 230\,GHz can be extrapolated assuming the same spectral index up to 230\,GHz and the turnover frequency in between  230\,GHz \citep[e.g.,][]{Yusef-zadeh_2006b, Bower_2015} to 1\,THz \citep{Bower_2019}. 
For the spectral index of \revnew{the} single power law, $\beta\sim0.39$, the extrapolated flux density at 230\,GHz is $\sim\,$2.73\,Jy. The lower and upper \revnew{limits} of the extrapolated flux densities are then estimated from the spectral indices for lower and higher frequency ranges (i.e., $\beta\sim\,$0.29 and 0.44), respectively. With this, the compact flux density of \sgra at 1.3\,mm in 2017 April, observed with the EHT, is expected to be $\sim2.73^{+0.23}_{-0.59}\,$Jy. 
\\

\section{Nonthermal electron emission models}
\label{sec:nonthermal_model}

The details of the physical nature of radio emission from \sgra \revnew{remain} elusive. 
Here we discuss the emission models and indicatives of the 
existence of nonthermal electrons to explain the wavelength-dependent intrinsic sizes of \sgra. 
Pioneering work by \citet{Ozel_2000} discussed a possible explanation of the prominent shoulder (excess) at $\lesssim100\,$GHz in the spectral energy distribution (SED) of \sgra by hybrid thermal-nonthermal electron population. The SED discussed in \citet{Ozel_2000}, however, was the assembly of non-simultaneously measured radio flux values and mostly observed by single dish data so that it potentially contained \revnew{a} significant amount of emissions originated from extended region around \sgra. 
Therefore, it is not entirely clear whether such a hybrid thermal-nonthermal electron population is really required or the observed excess is just a superposition of emissions from extended regions that cannot reflect intrinsic properties of \sgra.

In this work, we obtain the VLBI-measured quasi-simultaneous intrinsic sizes and the corresponding flux densities at 13, 7, and 3.5\,mm for the first time. Using these data, we discuss the necessity of nonthermal electron population in \sgra.
Note that the question about the dominant radio emission model of \sgra, either an accretion flow or a jet, is still open for debate. 
Since there is no clear observational evidence of jet eruption in \sgra on VLBI-scales yet, we will focus on the accretion flow model in this section. The possible jet model can be compared with the results from \citetalias{Issaoun_2019} 
\footnote{
Comparing the structure of \sgra at 3.5\,mm with the 3D general relativistic magnetohydrodynamic (GRMHD) simulations, \citetalias{Issaoun_2019} suggested the plausible models of 1) accretion flow dominated or 2) jet dominated with small viewing angles $\lesssim20^\circ$ (Figure~9 and 10 in their paper). 
With the same simulations, our results at 13 and 7\,mm disfavor the accretion flow dominated model (for both thermal and thermal/nonthermal hybrid cases). As for the jet dominated model, the small viewing angle is preferred, $\lesssim30^\circ$.}. 

We examine the theoretical images of radiatively inefficient accretion flow (RIAF) based on the Keplerian shell model \citep{Falcke_2000,Broderick_2006,Pu_2016,Kawashima_2019}. 
\rev{
The radiative transfer is solved using general relativistic, ray-tracing radiative transfer (GRRT) code, \texttt{RAIKOU} \citep{Kawashima_2019, Kawashima_2021a, Kawashima_2021b}. The cyclo-synchrotron via thermal electrons \citep{Mahadevan_1996} and synchrotron processes via nonthermal electrons \citep[see, e.g., ][]{Dexter_2016} are incorporated. For the sake of simplicity, the angle between the ray and magnetic field $\theta_{\rm B}$ is assumed to be $\sin \theta_{\rm B} = \pi/4$ for the calculation of emission/absorption via the nonthermal electrons, which corresponds to the average for \revnew{the} isotropic, random magnetic field. 
}

\begin{figure}[h]
\centering 
\includegraphics[width=0.95\linewidth]{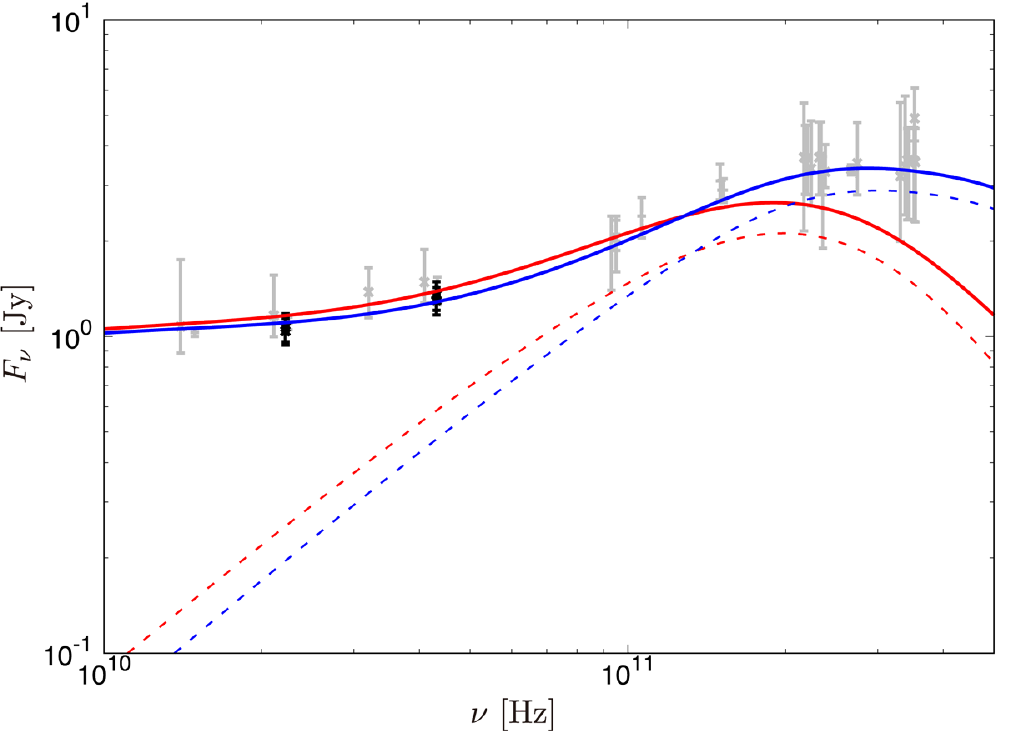}
\caption{
Theoretical SEDs for the models with  $i = 20^{\circ}$ (red) and $60^{\circ}$ (blue).
The thick-solid and thin-dashed lines represent the results taking into account thermal plus nonthermal electrons and thermal electrons only, respectively.
The black data points show the observation data at 22 and 43\,GHz in this work. The gray data points display the previous observational data \citep{Falcke_1998, Bower_2015}.
}
\label{fig:SED}
\end{figure}

We set the number density of thermal electrons $n_{\rm e,th}$, that of nonthermal electrons $n_{\rm e,nth}$,  electron temperature $T_{\rm e}$, and the magnetic field $B$ at $(r, \theta)$, where $r$ and $\theta$ is the spherical radius and the polar angle in the Schwarzschild coordinate, in the Keplerian shell model as follows: 

\vspace{-7pt}
\begin{equation}
\begin{aligned}
    n_{\rm e} = 6 {\times} 10^7 (r/r_{\rm g})^{-1.1} \exp (-z^2/2H^2) ~ [{\rm cm}^{-3}],\\ n_{\rm e, nth} = 2.5 {\times} 10^5 (r/r_{\rm g})^{-1.75} \exp (-z^2/2H^2) ~ [{\rm cm}^{-3}], \\
    T_{\rm e} = 1.5 {\times} 10^{11} (r/r_{\rm g})^{-8.4} ~ [{\rm K}], \\
    B^2/8\pi = 0.1 n_{\rm e,th}m_{\rm p}c^2r_{\rm g}/6r ~ [{\rm erg ~ cm}^{-3}]. \\  
\end{aligned}
\end{equation}
Here, $m_{\rm p}$ is the proton mass, $r_{\rm g} (= GM/c^2)$ is the gravitational radius, $G$, $M$ and $c$ are the gravitational constant, the BH mass, and the speed of light, respectively. 
We assume the $r_{g}$ with its mass $M = 4.1\times10^6 M_{\odot}$ and distance 8.1\,kpc. 
The scale height of the accretion flow is set to be $H/R = 0.5$, where $R$ is the cylindrical radius $R = r \sin \theta$. 
The acceleration mechanism of the electrons is uncertain, although they may be accelerated by the magneto-rotational instability (MRI) turbulence \citep{Jynn_2014}, magnetic reconnection  \citep{Hoshino_2013, Ball_2018, Werner_2018, Ripperda_2020} and/or by other mechanisms. 
Therefore we simply assume the energy spectrum of the nonthermal electrons to be a single power law distribution  $\propto {\gamma}^{-3.5}$ in the range $10^2 \le \gamma \le 10^6$, where $\gamma$ is the Lorentz factor of the electrons whose power-law index and the minimum Lorentz factor are same as those in \citet{Broderick_2011} and \citet{Pu_2016}. 
\rev{
Here, a slightly extended spatial distribution of nonthermal electrons $n_{\rm e, nth} \propto r^{-1.75}$ (in previous works, e.g., \citealt{Broderick_2011} and \citealt{Pu_2016}, $\propto r^{-2.02}$) is found from our observational data. 
Note that this is mainly due to the smaller uncertainty of our results (\autoref{fig:SED}) and the main conclusion of nonthermal electron population is not strongly affected by the small difference. 
}
The consequent higher fluxes at 1.3\,cm and 7\,mm require the spatially extended emission region due to the synchrotron emission via  nonthermal electrons. 
\rev{
Note that this is broadly applicable, not only valid for the parameter-set chosen above, where the Keplerian shell model holds. As is shown later, this can be confirmed by comparing the radius of the outer edge of the accretion flow and the observed size at each wavelength.}

\begin{figure*}[h!]
\centering 
\includegraphics[width=0.9\linewidth]{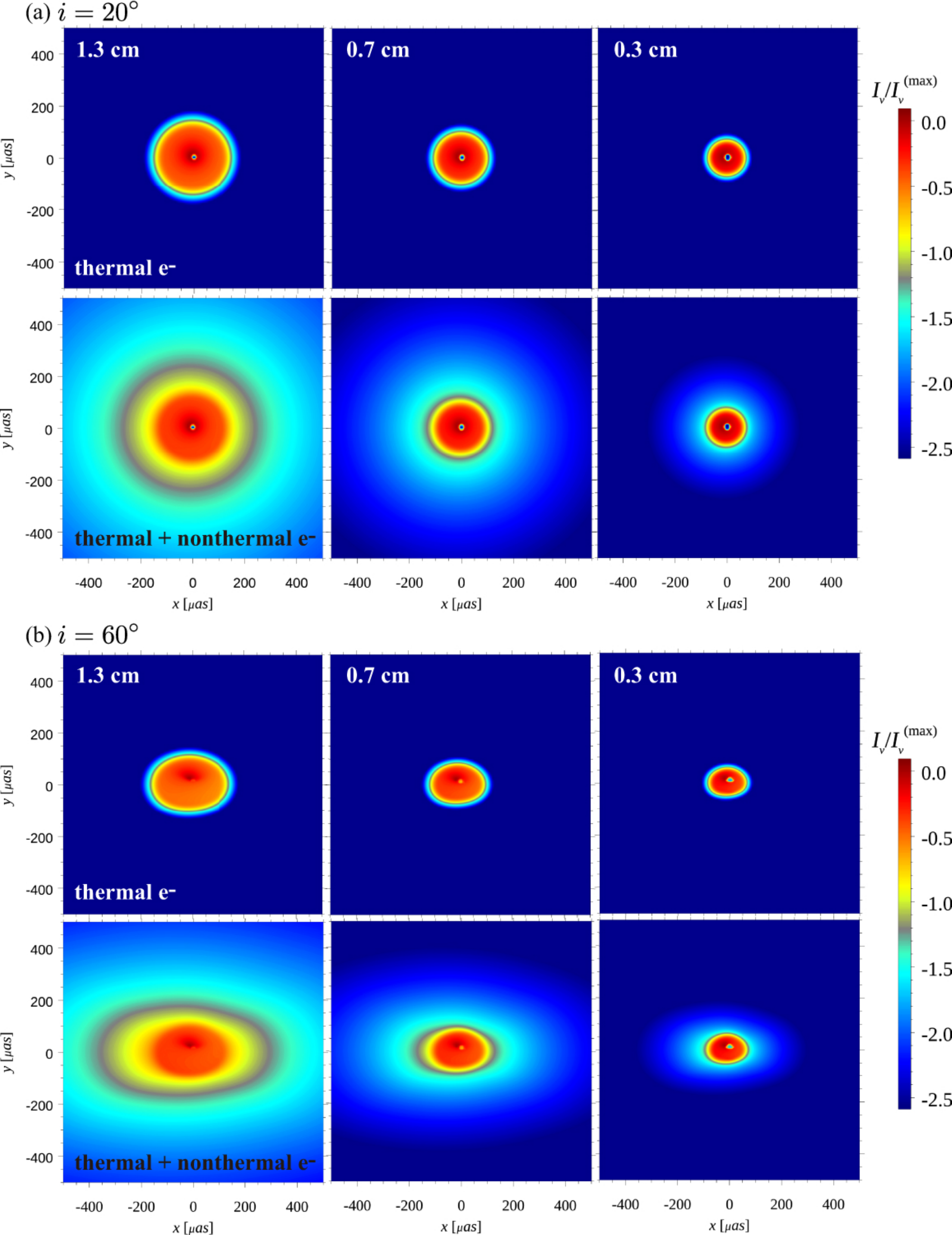}
\caption{
Theoretical images of accretion flow in Sgr A* observed at the wavelength 1.3 cm (left), 0.7 cm (middle), and 0.3 cm (right), based on the Keplerian shell model.
The viewing angle $i$ is (a) $20^{\circ}$ and (b) $60^{\circ}$.
The synchrotron emission/absorption processes are calculated taking into account the thermal electrons only (top) and both of the thermal and nonthermal electrons (bottom).}
\label{fig:simulation}
\end{figure*}

\autoref{fig:simulation} displays the resultant images of Sgr A* accretion flow.
The intensity is shown in the log scale. 
All of the models with only thermal electrons show the very compact core images. 
For example, at 1.3 cm, the intensity drastically decreases outside the diameter $\sim $ 300 $\mu$as, i.e., the intensity outside this diameter is lower than the order of magnitude of the peak intensity and this diameter is too small compared with the measured one.
This compact emission region is a consequence of the low synchrotron frequency in the outer region \rev{as explained in the next paragraph}.
On the other hand,  the synchrotron frequency can be higher than $\sim 10$\,GHz if the nonthermal electrons are assumed, so that the resultant images for the models with nonthermal electrons show \revnew{a} larger emission region whose diameter is $\sim ~500 ~\mu$as. 
This is qualitatively consistent with a pioneering theoretical work assuming one dimensional plasma structure \citep{Ozel_2000}. 
This indicates that our observed images provide the direct detection of the nonthermal electrons and \revnew{a} sufficient amount of nonthermal electrons should exist in the outer region, as have been proposed by the previous theoretical works focusing on the spatially unresolved SED \citep[e.g.,][]{Yuan_2003}\footnote{The importance of nonthermal electrons is also pointed out to explain the variability of the X-ray flux during \revnew{the} flaring state in Sgr A*, by carrying out GRMHD simulations with injecting nonthermal electrons and subsequent GRRT computations by taking into account synchrotron emission/absorption processes \citep{Ball_2016}.}. 
The image enlargement due to the synchrotron emission via the nonthermal electrons \revnew{is} also \revnew{consistent} with horizon-scale theoretical images at 230\,GHz \citep{Mao_2017, Chael_2017}. 
\autoref{fig:SED} shows a comparison between the observed flux densities and our theoretical SEDs. 
We find that the nonthermal synchrotron emission can well explain the overall excess component below 100\,GHz, including 22 and 43\,GHz observed by the EAVN. 

\begin{figure*}[t]
\centering 
\includegraphics[width=0.49\linewidth]{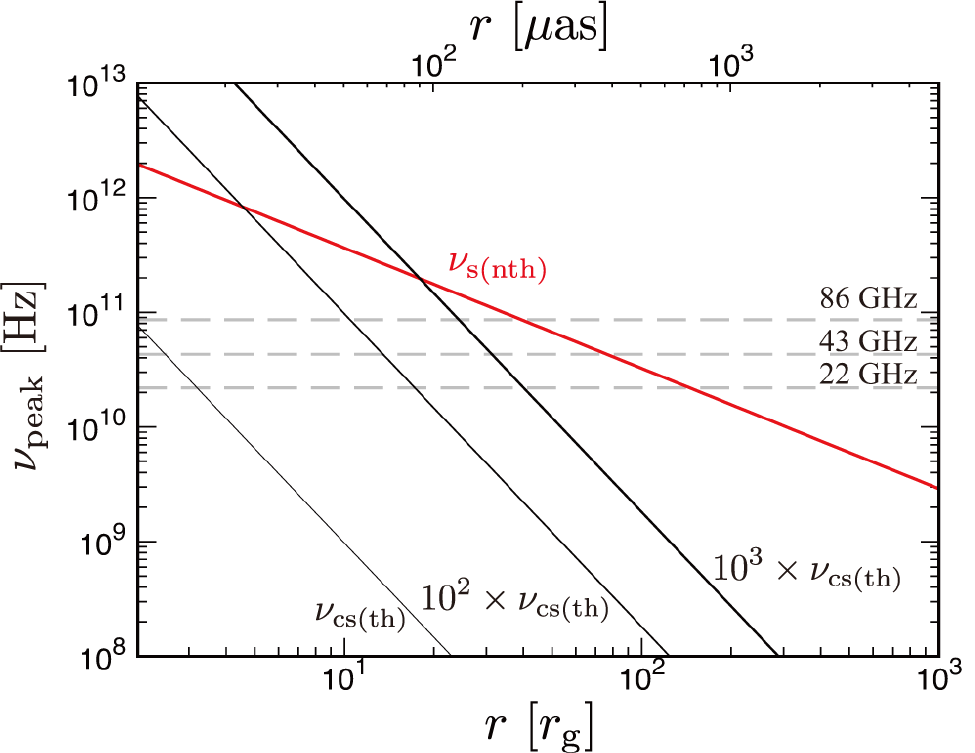}
\includegraphics[width=0.49\linewidth]{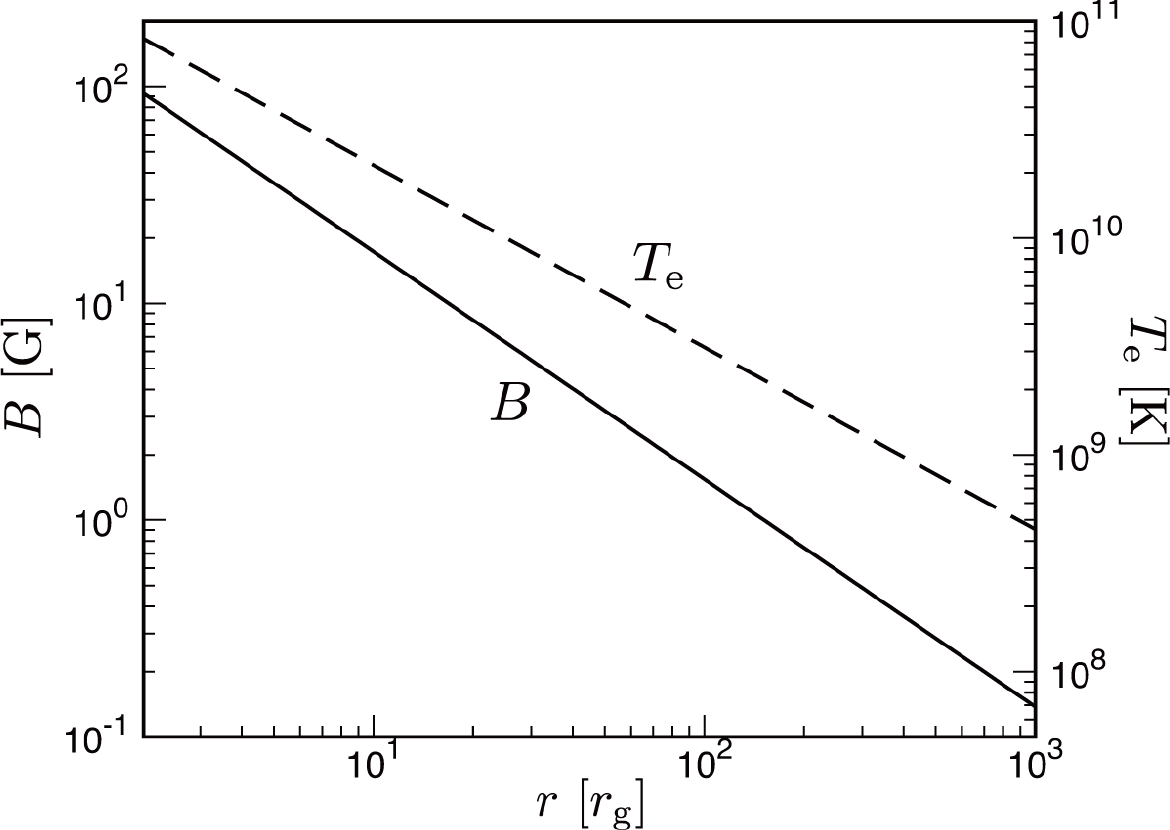}
\caption{
\rev{
(Left) Peak frequency of cyclo-synchrotron emission via thermal electrons (black) and synchrotron emission via nonthermal electrons (red).
(Right) Radial profile of magnetic field strength (solid line) and electron temperature (dashed line) of our Keplerian shell model.
}}
\label{fig:nu_peak_B_Te}
\end{figure*}

\rev{
We further interpret the image size of the theoretical models shown in Figure \ref{fig:simulation}, by considering the peak frequency of synchrotron and cyclo-synchrotron emissivity. 
The peak frequency of synchrotron and cyclo-synchrotron emissivity $\nu_{\rm peak}$, is generally expressed in terms of $B$, $T_{\rm e}$ (for thermal electrons), and minimum Lorentz factor $\gamma_{\rm e, min}$ (for nonthermal electrons with the power-law index greater than 2) in the accretion flow. 
With the Keplerian shell model, the $\nu_{\rm peak}$ can then be expressed as a function of the accretion flow radius. One can roughly estimate the outer edge of \revnew{the} bright core by the radius at which an observing frequency is equal to that of emissivity peak frequency, since the emission will become drastically faint if the observing frequency is greater than $\nu_{\rm peak}$. 
\autoref{fig:nu_peak_B_Te} (left) shows the peak frequency of cyclo-synchrotron emissivity via thermal electrons and \revnew{synchrotron} emissivity via nonthermal electrons, as a function of the emitting radius according to our Keplerian shell model. The corresponding spatial distributions of  $T_{\rm e}$ and $B$ are shown in \autoref{fig:nu_peak_B_Te} (right).
The peak frequency of our \revnew{nonthermal} synchrotron emissivity can be  described as $\nu_{\rm s (nth)} = \nu_{\rm c} \gamma_{\rm e, min}^2$ because of our steep power-law index $3.5 (> 2)$ of nonthermal electrons. 
Here $\nu_{\rm c} (= 3eB \sin \theta_B/ 4\pi m_{\rm e}c$) and $e$
are the cyclotron frequency of electrons and the elementary charge, respectively. 
As for the peak frequency of thermal cyclo-synchrotron emissivity, it is approximated as 
$10^{2} \nu_{\rm cs (th)} \lesssim \nu_{\rm peak} \lesssim 10^{3} \nu_{\rm cs (th)}$ for $T_{\rm e} \gtrsim 10^{10}$K 
where $\nu_{{\rm cs (th)}} = (3eB/4\pi m_{\rm e}c)\Theta_{\rm e}^2$ (see Figure~5 and 6 in \citealt{Mahadevan_1996}). 
In \autoref{fig:nu_peak_B_Te} (left), we find that the expected outer radius of the accretion flow seen at $22$ GHz for thermal cyclo-synchrotron case corresponds to only  $r \approx  20-40~r_{\rm g}$ ~ ($\lesssim 100 \mu {\rm as}$), 
while that of nonthermal synchrotron emitting accretion flow realizes $r \approx 200~r_{\rm g}$ ($\sim {\rm several} \times 100 \mu {\rm as} $) by the existence of nonthermal electrons in the outer region. 
The same logic can be applied to 43~GHz.
Here our model, $B \propto r^{-p_1}$ and $T_{\rm e} \propto r^{-p_2}$ with $p_1 \sim p_2 \sim 1$, is consistent with the recent simulation studies \citep[e.g.,][]{Chael_2017,Ressler_2020a}.
This strongly infers that an unusual disk model, whose electron temperature is high and magnetic field is strong in the outer disk region, is required to explain the observed image size of Sgr A* at 22 and 43 GHz if we do not assume nonthermal electrons.}
\footnote{We also examined the calculation of images assuming the strong magnetic field, which is more than an order of magnitude stronger than the standard magnetic field strength assumed in the Keplerian shell model, mimicking the magnetically arrested accretion flow \citep{Narayan_2003,Tchekhovskoy_2011,McKinney_2012,Narayan_2012}. However, the resultant image-size assuming thermal electrons is still small.}
\rev{Thus, we draw the conclusion that nonthermal electrons are necessary to explain the observed sizes of \sgra measured at 22 and 43\,GHz if the disk-dominant emission is assumed.}
\footnote{\rev{At the same time, it is fair to state the caveat that we neglect a possible case of a jet. If this exists in \sgra, a larger emission size might be reproduced by \revnew{a} pure thermal electron model \citep[e.g.,][]{Moscibrodzka_2014}. Further observational and theoretical studies of jets are therefore needed to draw a final conclusion. }}

Lastly, we discuss the dependence of the axial ratio on the viewing angle $i$ of observers. 
Since an extensive and detailed parameter survey of $i$ is beyond the scope of this study, here we just investigate two representative 
cases: a small viewing angle \citep[e.g.,][]{Gravity_2018b} and large viewing angle \citep[e.g.,][]{Markoff_2007, Broderick_2011}. 
Here we set $i=20^{\circ}$ for the small viewing angle case, while $i=60^{\circ}$ is chosen for the large viewing angle case.
From the model predicted images shown in \autoref{fig:simulation}, we find that the morphology of the emission region is almost isotropic for $i=20^{\circ}$, while the axial ratio is $\sim2$ for $i=60^{\circ}$.
Therefore, the axial ratio $\simeq1.2-1.3$ from our measurements may suggest that a small viewing angle is preferred to a large viewing angle. 
Note however that the discussion here is just one heuristic argument and further investigations are needed to reach a final conclusion. 
We also note that the change of the viewing angle possibly occurs in the near future, as a consequence of the tilt of accretion flow due to the misalignment of the directions of magnetic flux and averaged angular momentum of the accreting gas fed by the Wolf-Rayet stars \citep{Ressler_2020a, Ressler_2020b}. 
The successive observations are therefore needed to study the exact viewing angle of the accretion flow and its possible variation in time. 
\\

\section{Summary} \label{sec:summary}

In this study, we present the intrinsic properties of \sgra at 1.3\,cm and 7\,mm from the EAVN observations. 
Through the imaging and Gaussian model fitting, the scattered size of \sgra in \revnew{an} ensemble-average limit is first derived which is dominated by the scattering effect. 
Adopting the recent scattering kernel model, the self-calibrated visibilities and closure amplitudes are deblurred. The unscattered structure of \sgra is then obtained from the Gaussian model fitting onto the deblurred data. As a result, we find a single, symmetric Gaussian model well explains the structure.  
From the closure amplitudes, the major axis sizes are $\sim704\,\mu$as (axial ratio$\sim1.19$, PA$\sim95^\circ$) at 1.3\,cm and $\sim300\,\mu$as (axial ratio$\sim1.28$, PA$\sim95^\circ$) at 7\,mm. 
Together with the 3.5\,mm size which has been quasi-simultaneously measured \citepalias{Issaoun_2019}, the wavelength dependent source size is found with the power law index $\epsilon=1.2\pm0.2$. 

The expected size of \sgra at 1.3\,mm is extrapolated to \rev{$36^{+18}_{-16}\,\mu$as and $27^{+13}_{-12}\,\mu$as} toward the major and minor axis, respectively. 
From the total flux densities at three wavelengths, in addition, the spectral index is derived as $\beta\sim0.39$ and the extrapolated compact flux density at 1.3\,mm is $\sim2.73$\,Jy. 
With more (quasi) simultaneous observations at broader wavelengths region, the wavelength dependence may be constrained more robustly and the long-term time variation of structure can be further investigated. 

As for the dominant emission scenario, we have compared the measured intrinsic size of \sgra with the accretion flow dominated model, especially the RIAF based on the Keplerian shell model. 
In this case, the intrinsic sizes at both wavelengths are a factor of a few larger than those predicted with purely thermal electron distribution. We find that this size-mismatch problem can be solved by including nonthermal electron components. 
The obtained axial ratio which is almost isotropic also indicates the small viewing angle of \sgra, $\lesssim30-40^\circ$. 
This is consistent with the previous study of measuring the rotating hot spot \citep{Gravity_2018b} and the 3D GRMHD simulations with the jet-dominated model \citepalias{Issaoun_2019}. 
To discriminate each scenario, additional multi-wavelength observations will be of great help, for instance investigating the frequency-dependent radio core position shifts (I.~Cho et al. in prep.). 
\\

\acknowledgments
\rev{
We thank the anonymous referee and the EHT Collaboration internal reviewers \& Publication Committee, G.C. Bower, L. Loinard, and K. Rygl, who reviewed the manuscript.} 
We are grateful to all staff members in EAVN who helped to operate the array and to correlate the data. The KVN is a facility operated by Korea Astronomy and Space Science Institute (KASI) and VERA is a facility operated by the National Astronomical Observatory of Japan (NAOJ) in collaboration with associated universities in Japan. 
TIA is operated by Shanghai Astronomical Observatory.
UR is operated by Xinjiang Astronomical Observatory. 
HT is operated by NAOJ and Ibaraki University. 
The KVN operations are supported by Korea Research Environment Open NETwork (KREONET) which is managed and operated by Korea Institute of Science and Technology Information (KISTI). 
The theoretical calculations were carried out on the XC50 at the Center for  Computational Astrophysics, National Astronomical Observatory of Japan. 
I.C. and J.P. are supported by the National Research Foundation of Korea (NRF) via a Global PhD Fellowship Grant (NRF-2015H1A2A1033752 and NRF-2014H1A2A1018695, respectively). 
\rev{I.C. acknowledges financial support 
by the Spanish Ministerio de Econom\'{\i}a y Competitividad (grants PID2019-108995GB-C21).
} 
G.-Y.Z. and T.J. are supported by Korea Research Fellowship Program through the NRF (NRF-2015H1D3A1066561). 
G.-Y.Z. acknowledges financial support from the State Agency for Research of the Spanish MCIU through the ``Center of Excellence Severo Ochoa'' award to the Instituto de Astrof\'{\i}sica de Andaluc\'{\i}a (SEV-2017-0709). 
K.A. is supported by the National Science Foundation (NSF) through grants AST-1440254, AST-1614868, and AST-2034306. 
This work is partially supported by JSPS KAKENHI Grant Numbers JP18K03656 (M.K.), JP18H03721 (K.N., M.K., K.H.), JP18K13594 (T.K.), JP18KK0090 (K.H.), JP19H01943 (K.H.), JP21H01137 (K.H., M.K.), and the Mitsubishi Foundation grant No. 201911019 (K.H.). 
J.-C.A acknowledges support from the Malaysian Fundamental Research Grant Scheme (FRGS) FRGS/1/2019/STG02 /UM/02/6.  
M.D.J. acknowledges support from the National Science Foundation (AST-1716536, AST-1935980) and the Gordon and Betty Moore Foundation (GBMF-5278). 
X.-P.C. and B.-W.S. are supported by Brain Pool Program through the National Research Foundation of Korea (NRF) funded by the Ministry of Science and ICT (2019H1D3A1A01102564). 
R.-S.L. is supported by the Max Planck Partner Group of the MPG and the CAS and acknowledges the support by the Key Program of the National Natural Science Foundation of China (grant No. 11933007) and Research Program of Fundamental and Frontier Sciences, CAS (grant No. ZDBS-LY-SLH011). 
J.P. acknowledges financial support through the EACOA Fellowship awarded by the East Asia Core Observatories Association, which consists of the Academia Sinica Institute of Astronomy and Astrophysics, the National Astronomical Observatory of Japan, Center for Astronomical Mega-Science, Chinese Academy of Sciences, and the Korea Astronomy and Space Science Institute. 
S.T. acknowledges financial support from the NRF via Basic Research Grant 2019R1F1A1059721. 
This work was also supported in part by MEXT SPIRE, MEXT as ``Priority Issue on post-K computer'' (Elucidation of the Fundamental Laws and Evolution of the Universe) and as ``Program for Promoting Researches on the Supercomputer Fugaku'' (High-energy astrophysical phenomena in black holes and supernovae). 

$Software$: 
{\tt AIPS} \citep{Greisen_2003}, 
{\tt Astropy} \citep{Astropy}, 
{\tt DIFMAP} \citep{Shepherd_1994}, 
{\tt eht-imaging} \citep{Chael_2018}, 
{\tt Matplotlib} \citep{Matplotlib},
{\tt NumPy} \citep{Numpy}, 
{\tt PyMC} \citep{Pymc3}, 
{\tt SciPy} \citep{Scipy}, 
{\tt SMILI} \citep{Akiyama_2017a, Akiyama_2017b, Smili_2019},

\begin{appendix}

\section{SMILI imaging} \label{app:smili_imaging}
\rev{
SMILI imaging is one of the regularized maximum likelihood (RML) methods. 
This finds an image ($I$) which minimizes a specified objective function, 
\begin{align}
 \label{eq::objfunc}
 J(I) = \sum_{\mathclap{\text{data terms}}} \alpha_{\textrm{D}} \chi^2_{\textrm{D}}\left(I\right) - \sum_{\mathclap{\text{regularizers}}} \beta_{\textrm{R}} S_{\textrm{R}}\left(I\right),
\end{align}
where the first and second term corresponds to a measure of the inconsistency of the image (e.g., goodness-of-fit functions, $\chi^2_D$) and regularization ($S_R$), respectively. 
These terms often have opposite preferences for a fiducial image, so their relative impact in the minimization process is specified with the coefficients of regularization terms (i.e., hyperparameters, $\alpha_R$ and $\beta_R$). 
Regularizers in SMILI are explored to constrain the image characteristics, for instance sparsity ($\ell_1$ norm) and smoothness (total variation and total squared variation), and both of them can be simultaneously favored in the minimization of the objective function. 
Note that the fiducial values of regularizers can be different for different observational data towards \revnew{the} same target source, since they are found by minimizing the inconsistency term at the same time. 
For more details of regularizer definitions, see Appendix\,A of \citet{ehtc2019d}. 
}

\section{Error estimates} \label{app:error}

\begin{figure*}[h]
\centering 
\includegraphics[width=\linewidth]{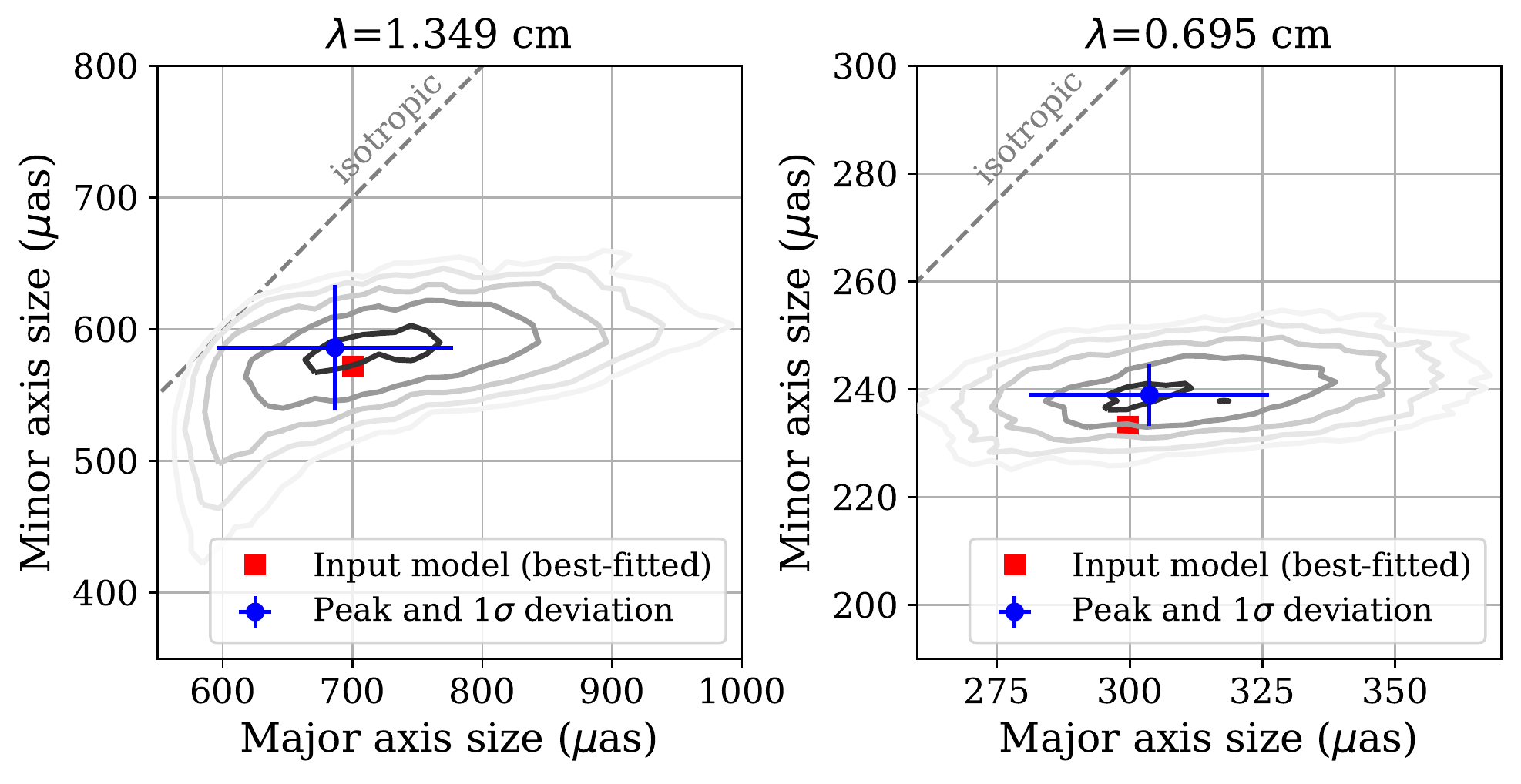} \\
\caption{
Two-dimensional histogram density of the derived intrinsic sizes of \sgra by deconvolving the random phase screen: 1.3 (left) and 0.7\,cm (right). 
The input model (i.e., best-fitted model) is first given (red square) and \revnew{convolved} with the random phase screen. After that, deblur the scattering kernel and fit a single Gaussian to estimate the sizes. 
The same procedure is repeated with $\sim150$ number of different scattering parameters and 1,000 times of random phase screen for each scattering parameter (gray contours; 5, 10, 20, 40, 80\,\% of the peak number are shown). 
As a result, the possible deviation of the size measurement due to the scattering effects is derived (blue circle with error bar). Note that no significant biases are found from the peak value.} 
\label{fig:result_synthetic}
\end{figure*}

The size uncertainties of \sgra are estimated with the goodness-of-fit from Monte Carlo method ($\sigma_{\rm fit}$) and the stochastic random phase screen within the error range of scattering parameters ($\sigma_{\rm scat}$) (\autoref{tab:err_comp}). 
To test the latter effects, the best fitted Gaussian models are scattered with the random phase screen using the {\tt eht-imaging} library and its visibilities are generated with corresponding observations. Then the same procedures of elliptical Gaussian model fitting for an ensemble-average image, scattering kernel deblurring, and intrinsic size fitting with a single Gaussian model are applied. 
While the fixed scattering parameters, $\alpha=1.38$ and $r_{\rm in}=800\,$km, are used for ensemble-average image, different scattering parameters within its possible ranges (e.g., $\alpha=1.38^{+0.08}_{-0.04}$, $r_{\rm in}=800\,\pm\,200\,$km; \citetalias{Johnson_2018}) have been tested with randomly selected $\sim150$ numbers for \revnew{the} intrinsic size of \sgra. 
To derive reasonable deviations, these processes are repeated 1,000 times for each scattering parameter. 
As a result, the standard deviations of the sizes are larger towards major axis with a consistent position angle with the best-fitted model (see \autoref{fig:result_synthetic}). 
Note that the peak intrinsic size is well consistent with the input model (i.e., the best-fitted Gaussian) so that no systematic biases are found. 
\\

\begin{table}[ht]
\centering
\addtolength{\tabcolsep}{-2pt}
\small
\caption{Size uncertainties}
\begin{tabular}{lccccc}
\hline
\hline
$\lambda$ (cm) & 
Axis & 
$\sigma^{\rm en}_{\rm fit}$ ($\mu$as) & 
$\sigma^{\rm en}_{\rm scat}$ ($\mu$as) & 
$\sigma^{\rm int}_{\rm fit}$ ($\mu$as) & 
$\sigma^{\rm int}_{\rm scat}$ ($\mu$as) 
\\
\hline
1.349 & Major & $\pm\,$10.8 & $\pm\,$25.7 & $\pm\,$45.9 & $\pm\,$91.1 \\
      & Minor & $\pm\,$29.2 & $\pm\,$7.4 & $^{+61.8}_{-70.1}$ & $\pm\,$47.8 \\
0.695 & Major & $\pm\,$3.4 & $\pm\,$8.3 & $\pm\,$10.2 & $\pm\,$22.5 \\
      & Minor & $\pm\,$19.6 & $\pm\,$2.4 & $^{+31.8}_{-34.2}$ & $\pm\,$5.8 \\
\hline
\end{tabular}
\\ \vspace{0.3cm}
\raggedright{\textbf{Note. }
From left to right, observing wavelength, \revnew{the} axis of Gaussian structure, \revnew{the} uncertainty of ensemble-average image size (from fitting and stochastic random phase screen), and intrinsic image size (from fitting and overall scattering effects). 
The Monte Carlo fitting error is slightly different for each measurement (i.e., Gfit/Amp, SMILI/Amp, and CA), and here we show the uncertainty from CA. The overall uncertainty of size measurement is obtained by the quadratic sum of each error component (\autoref{tab:results}).}  
\label{tab:err_comp}
\end{table}

\section{Intensity map of the Keplerian shell model in one-\revnew{dimensional} slice of the observer screen} \label{app:1D_slice_model}

We present the one-dimensional (1D) slices of the intensity map shown in \autoref{fig:simulation}. 
Since it is uncertain what is the true peak-intensity in the observed blurred images due to the limited beam size, 
we qualitatively discuss the theoretical intensity map. 

\begin{figure*}[h]
\centering 
\includegraphics[width=0.96\linewidth]{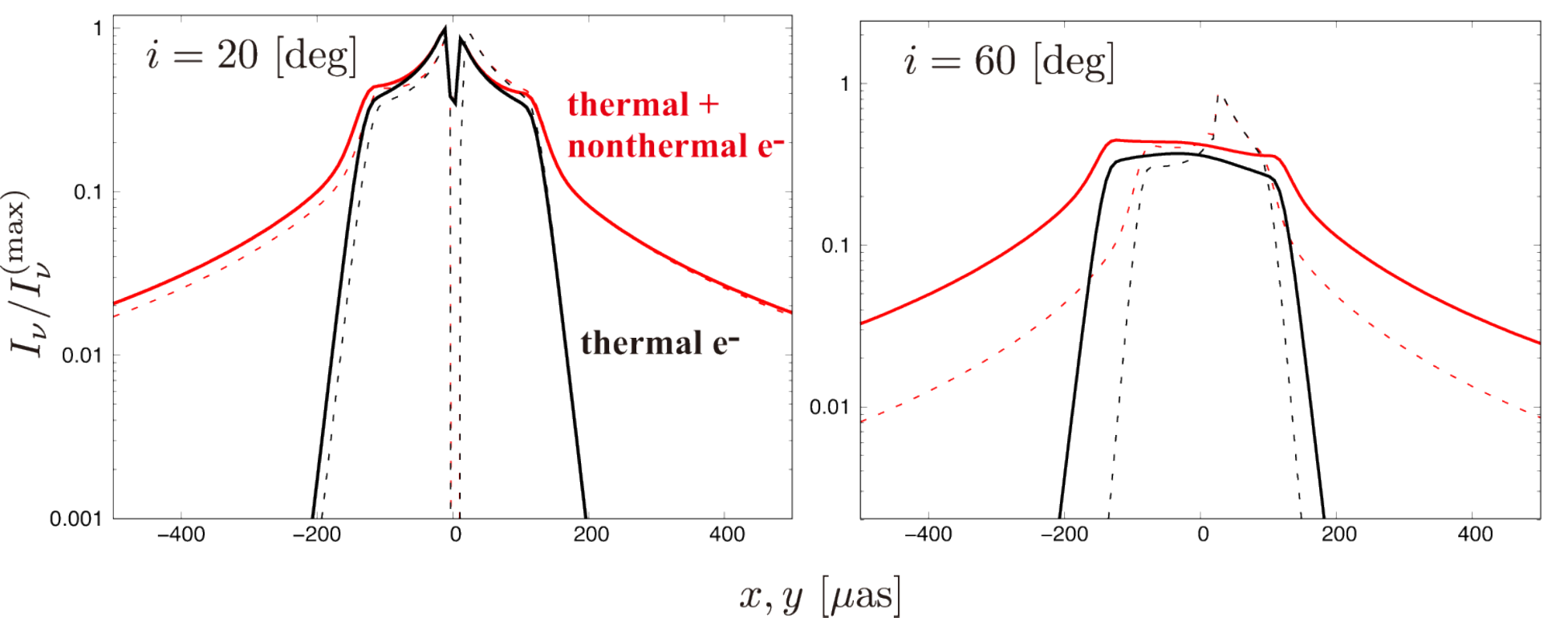} \\
\caption{
One-dimensional slice of the intensity map of theoretical image shown in \autoref{fig:simulation} with the viewing angle $i = 20^{\circ}$ (left) and $60^{\circ}$ (right). The photon frequency, $\nu$, is 22\,GHz. The solid and dashed lines show the distribution of intensity in $x$ and $y$ direction at $y =0$ and $x=0$, respectively. The results for the model with thermal and nonthermal electrons (red) and with thermal electrons only (black) are displayed.
The intensity is normalized by the maximum value in the $x$-$y$ plane of the screen (i.e., by the peak value in the entire region of the screen).
} 
\label{fig:1D_slice_model}
\end{figure*}

\autoref{fig:1D_slice_model} displays the 1D intensity map in the observer screen.
For the model with $i=20^{\circ}$, the 1D intensity profile in the $x$-direction at $y=0$ (solid line) is almost identical to that in $y$-direction at $x=0$ (dashed line), i.e., the nearly isotropic emission feature can be confirmed in this figure.
On the other hand, for the model with $i=60^{\circ}$, the 1D intensity profile is significantly \revnew{anisotropic}, because of the finite scale height of the accretion flow. 

The sharp intensity peak appeared in the model with $i=20^{\circ}$ due to the emission from the inner part of the disk, while it disappeared for the model with $i = 60^{\circ}$ in the horizontal direction, because the photons emitted from the  inner accretion flows are obscured by the outer accretion flow (i.e., the self-occultation effect).
We also note that the peak intensity for the model with $i=60^{\circ}$ appeared in $y > 0$ rather than $y<0$ because of the more significant effect of the self-occultation  in $y < 0$ (observer side) than in $y > 0$ (counter side). 

\end{appendix}

\bibliography{Cho_2021_ApJ}{}
\bibliographystyle{aasjournal}

%


\end{document}